\documentclass[11pt, a4paper]{article}
\pdfoutput=1

\usepackage{amsmath}
\usepackage{amsfonts}
\usepackage{amssymb}
\usepackage{graphicx}
\usepackage{mathrsfs}
\usepackage{latexsym}
\usepackage{color}
\usepackage{cite}
\usepackage{slashed, cancel}
\usepackage{hyperref}
\usepackage{datetime}

\numberwithin{equation}{section}

\hypersetup{colorlinks, citecolor=bluscuro, linkcolor=bluscuro, urlcolor=bluscuro}
\definecolor{rossos}{rgb}{0.8,0.2,0.3}
\definecolor{bluscuro}{rgb}{0.15, 0.2, .85}
\definecolor{bluchiaro}{cmyk}{1,.3,0.,0.1}

\makeatother   
\newcommand{\GeV}{{\rm \,GeV}}
\newcommand{\TeV}{{\rm \,TeV}}

\def\de{\textrm{d}}

 \def\be   {\begin{equation}}   \def\ee   {\end{equation}}
 \def\ba   {\begin{array}}      \def\ea   {\end{array}}
 \def\bea  {\begin{eqnarray}}   \def\eea  {\end{eqnarray}}
 \def\bean {\begin{eqnarray*}}  \def\eean {\end{eqnarray*}}
 \def\nn{\nonumber}

\begin{document}

%
\begin{flushright} 
CERN-PH-TH/2013-151\\
SISSA  29/2013/FISI
\end{flushright}

\vspace{0.5cm}
\begin{center}

{\LARGE \textbf {
On the Validity of the Effective Field Theory
\\[0.3cm]
 for Dark Matter Searches at the LHC
}}
\\ [1.5cm]

{\large
\textsc{Giorgio Busoni}$^{\rm a,}$\footnote{\texttt{giorgio.busoni@sissa.it}},
\textsc{Andrea De Simone}$^{\rm a, b,}$\footnote{\texttt{andrea.desimone@sissa.it}},
}
\\[0.2cm]
{\large
\large \textsc{Enrico Morgante}$^{\rm c,}$\footnote{\texttt{enrico.morgante@unige.ch}},
\textsc{Antonio Riotto}$^{\rm c,}$\footnote{\texttt{antonio.riotto@unige.ch}}
}
\\[1cm]

\large{
$^{\rm a}$ 
\textit{SISSA and INFN, Sezione di Trieste, via Bonomea 265, I-34136 Trieste, Italy}

\vspace{1.5mm}
$^{\rm b}$ 
\textit{CERN, Theory Division, CH-1211 Geneva 23, Switzerland}

\vspace{1.5mm}
$^{c}$ 
\textit{D\'epartement de Physique Th\'eorique and Centre for Astroparticle Physics (CAP),\\
24 quai E. Ansermet, CH-1211 Geneva, Switzerland}
}
\end{center}

\vspace{0.5cm}

\begin{center}
\textbf{Abstract}
\begin{quote}
We discuss the limitations to the use of the effective field theory approach
to study dark matter at the LHC. We introduce and study a few  quantities, some of them independent of the ultraviolet completion of the dark matter
 theory, 
which quantify   the error made when using 
effective operators  to describe processes with very high momentum transfer. 
Our criteria indicate up to what cutoff energy scale, and with what precision, 
the  effective  description is valid,
depending on the dark matter mass and couplings.
\end{quote}
\end{center}

\def\thefootnote{\arabic{footnote}}
\setcounter{footnote}{0}
\pagestyle{empty}

\newpage
\pagestyle{plain}
\setcounter{page}{1}

\section{Introduction}
Identifying the nature and properties of  Dark Matter (DM), whose contribution to the total energy density budget of the universe amounts to about 30\% \cite{planck},   will have profound consequences both in high energy particle physics and in cosmology. A very intense  experimental activity is nowadays devoted to the search
for DM:  in  the so-called indirect searches of  DM  the goal is to  detect the products of 
DM annihilations or decays around the Milky Way \cite{review}; in  direct searches the hope is to 
detect the    scattering between DM and  heavy nucleons \cite{reviewdirect}; finally in collider searches at the Large Hadron Collider
(LHC)  the mono-jet \cite{monojetATLAS1, monojetCMS1, monojetATLAS2, monojetCMS2}
and mono-photon \cite{monogammaATLAS1, monogammaCMS1, monogammaATLAS2, monogammaCMS2} searches  are currently  under way 
to look for an indirect signature of DM production (for alternative kinds of DM searches
at the LHC see e.g.~Refs.~\cite{DeSimone:2010tf, Bai:2011wy, Fox:2012ee, Cotta:2012nj, Bai:2012xg, Bell:2012rg}).
The complementary interplay
of these different approaches  can considerably improve the discovery potential.

Despite the fact that the nature of the DM is not known, from the theoretical point of view the most studied 
candidate  is represented by a Weakly Interacting Massive Particle (WIMP): a neutral particle with weak-scale mass and weak interactions,  whose thermal relic density may naturally fit the observed DM  abundance. Given the plethora  of particle physics model beyond the SM providing such a candidate, it is  highly desirable to study the signatures of this DM candidate in a model-independent way. 
One possible  starting point is the  Effective Field Theory (EFT) approach where the interactions between the DM particle and the SM sector are parametrized by a set of effective (non-renormalizable) operators, generated after integrating out heavy mediators \cite{Beltran:2010ww, goo, Bai:2010hh, Goodman:2010ku, Rajaraman:2011wf, fox, Dreiner:2012xm, Chae:2012bq, ds, Dreiner:2013vla, Chen:2013gya}.
Since direct and indirect detection of WIMPs, as well as WIMP production at the LHC, all require an interaction of the WIMPs with 
the SM particles, and such an interaction may be generated by the same operator, the EFT approach has the advantage of facilitating the analysis of the 
 correlations between the various kinds  of experiments. Currently, the EFT approach is being used by the LHC collaborations in their DM searches
 in the  mono-jet/mono-photon plus missing energy channels. 

The EFT description is only justified whenever there is a clear separation between the energy scale of the process to describe 
and the scale of the underlying microscopic  interactions. 
 In  other contexts of  DM searches, the energy scales involved are such that the EFT
expansion is completely justified. For instance, for indirect DM searches 
the annihilations of non-relativistic DM particles 
in the galaxy occurs with momentum transfers of the order of the DM mass $m_{\rm DM}$;
in direct searches,  the momentum transfers involved in the scattering of DM particles with heavy nuclei are of the order of tens of keV.
In all these cases, it is possible to carry out an effective description in terms of operators with an Ultra-Violet (UV)  cutoff 
larger than the typical momentum transfer, and reliable limits on the operator scale can be derived   
(see e.g. Refs.~\cite{eftconstraints, eftdirectdetection}).

However, the situation is dramatically different at the LHC environment, 
where the energies involved can be very high, and the processes one wishes to describe
in terms of effective operators can actually occur at an energy beyond the validity of the
EFT itself. 
The effective non-renormalizable operators mimic the effect of heavy particles. Of course, 
such a description will be incorrect in the case that near future LHC searches provide a signal of a direct production of the heavy mediators. Supposing that this will not be the case, 
 one is left with the conclusion  that the EFT is valid as long as the energy scale of the process involving the DM and the SM particles is small compared to the energy scale associated to the heavy mediator. 
Thus, in this situation one should make a careful use of the EFT, making sure to use it
consistently and within its range of applicability.

 The goal of this paper is to answer a simple, yet basic, question: under what
 circumstances is the EFT approach reliable
for DM searches at the  LHC? We will introduce some  quantities to assess the validity of the EFT. Some of them  have the virtue of being independent of the UV completion of the DM theory and therefore can be adopted in their full generality. 
 The paper is organized as follows. In section \ref{sec:general}, we discuss some general 
 issues about the EFT and introduce a simple model with a heavy scalar mediator
 which will be a useful example to illustrate the points we make. 
 In section  \ref{sec:estimate}, we provide an estimate of the energy transfer which gives us the first indication
 of the goodness of the EFT. In section \ref{sec:validity}, we introduce various  model-independent ratios to quantify the validity of the 
EFT approach; in the same section we go back to the heavy scalar mediator model to 
analyze the EFT for a well-defined specific case. Finally, 
in section \ref{sec:conclusions} we draw our conclusions.

\section{General considerations}
\label{sec:general}

An EFT is a powerful and economical way to describe physical processes occurring
at a given energy scale in terms of a  tower of interactions, involving only 
the degrees of freedom present at such scale.
These interactions are generically non-renormalizable and with mass dimensions $\Lambda^k$, for
some $k\geq 1$.
For example, if one imagines that the UV
theory contains a heavy particle of mass $M$,
the low-energy effective theory at energies less than $\Lambda\sim M$  only contains the degrees of freedom lighter than $\Lambda$.
The effects of the heavy field in the processes at low momentum  transfer $Q_{\rm tr}\ll\Lambda$
are encoded by a series of interactions, scaling as $(Q_{\rm tr}/\Lambda)^k$ and whose coefficients are matched to reproduce the UV theory at $Q_{\rm tr}=\Lambda$.
Therefore, the scale $\Lambda$ sets the maximum energy at which the operator
expansion in the EFT can be trusted.
So generally speaking, the condition for the validity of an EFT is that the momentum transfer $Q_{\rm tr}$ in the relevant process one wants to describe  must be less than the energy scale 
$\Lambda$.

 In order to assess to what extent the effective description is valid, one has to compare
the momentum transfer $Q_{\rm tr}$ of the process of interest (e.g. 
$ p p\to \chi\chi$+jet/$\gamma$) to the energy scale $\Lambda$ and impose that
\be
\Lambda \gtrsim Q_{\rm tr}\,.
\label{condition}
\ee
Of course, there is some degree of arbitrariness in this choice, as one does not expect
a sharp transition between a valid and an invalid EFT, but more precisely that the observables
computed within the EFT are a less and less accurate approximation of the ones of the unknown UV
theory as  the cutoff scale $\Lambda$ is approached.
A more precise information on what $\Lambda$ is can only come from
knowing the details of the UV theory, the mass spectrum and the strength
of the interactions.

The lower limits on $\Lambda$, extracted from interpreting the experimental data
in terms of  effective operators,
should be considered together with the  condition (\ref{condition}) on the validity itself of the effective approach.
This means that  one has to make sure that the lower limits on $\Lambda$ 
 obtained by the experiments satisfy to some extent the condition (\ref{condition}).

\begin{figure}[t!]
\centering
\includegraphics[scale=0.3]{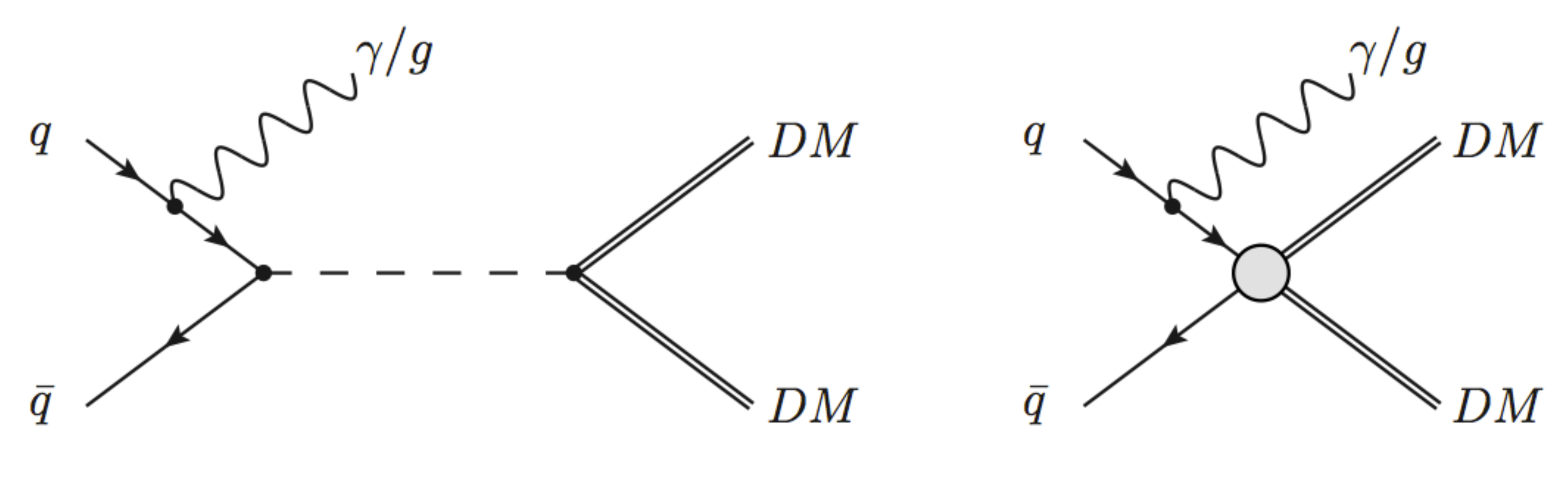}
\caption{\emph{{\small
The Feynman diagrams for DM pair production with ISR of a photon or jet,
for a model with scalar exchange \emph{(left panel)} and its effective
operator \emph{(right panel)}. We omitted the diagrams where the radiation
is emitted from the anti-quark.
 }}} 
\label{fig:diagrams}
\end{figure}

In order to clarify this point better,  let us consider a simple example, which will be the \textit{leitmotiv}
throughout this work. 
Let us consider DM to be a fermion, whose interactions with quarks are mediated by
a heavy scalar particle $S$ through the Lagrangian
\be
\label{lagr}
\mathscr{L}_{\rm UV}\supset \frac{1}{2}M^2 S^2
-g_q \bar q q S
-g_\chi \bar \chi \chi S\, .
\ee
At energies much smaller than $M$ the heavy mediator $S$ can be integrated out,
resulting in a tower of non-renormalizable operators for the fermionic DM interactions with quarks. 
The lowest-dimensional operator has dimension six
\be
\label{os1}
\mathcal{O}_S=\frac{1}{\Lambda^2}(\bar\chi\chi)
(\bar q q)\, ,
\ee
and the matching condition implies
\be
\frac{1}{\Lambda^2}=\frac{g_\chi g_q}{M^2}\, .
\label{Os}
\ee
The Feynman diagrams for the processes under consideration are 
depicted in Fig.~\ref{fig:diagrams}.
The processes where
a quark-jet is emitted from an initial gluon also contribute to the signal, but are suppressed
by a factor of about 4 at 8 TeV LHC with respect to the gluon emission, and for simplicity 
we will not consider them in this paper.
The procedure  of integrating out the heavy mediator and retaining the operator of lowest dimension
 can be viewed in terms of the expansion of the heavy particle propagator
\be
\label{ki}
\frac{1}{Q_{\rm tr}^2-M^2}
=-\frac{1}{M^2}\left(1+\frac{Q_{\rm tr}^2}{M^2}+\mathcal{O}\left(\frac{Q_{\rm tr}^4}{M^4}\right)\right)\,,
\ee
where only the leading  term $1/M^2$ is kept.
The higher-order terms in the expansion correspond to higher-dimensional operators.
It is obvious that retaining only the lowest-dimensional operator
is a good approximation as long as $Q^2_{\rm tr}\ll M^2\sim \Lambda^2$. 
Thus, the parameter $Q_{\rm tr}/M$ characterizes
the goodness of the truncation of the tower of effective  operators to the lowest
dimensional ones.

For the couplings to stay in the perturbative regime, one needs $g_q, g_\chi<4\pi$
(see Ref.~\cite{Shoemaker:2011vi} for an alternative criterion based on unitarity). Also, 
we need a mediator heavier than the DM particle $m_{\rm DM}$, that is 
 $M>m_{\rm DM}$. So, Eq.~(\ref{Os}) gives ~\cite{Goodman:2010ku}
\be
\Lambda\gtrsim \frac{m_{\rm DM}}{4\pi}\,,
\label{mover2pi}
\ee
which   depends linearly on the DM mass.
This is a very minimal requirement on $\Lambda$ and it is what, for instance,  ATLAS
uses in  Ref.~\cite{monojetATLAS2}.
On top of this condition, the validity of the truncation to the lowest order in the expansion (\ref{ki}) 
requires that  $Q_{\rm tr}<M$, i.e.
$ Q_{\rm tr}<\sqrt{g_q\,g_\chi} \Lambda < 4\pi \Lambda$,
so that 
\be
\label{4pi}
\Lambda>\frac{Q_{\rm tr}}{\sqrt{g_q g_\chi}}>\frac{Q_{\rm tr}}{4\pi}\,,
\ee
which depends on $m_{\rm DM}$ through $Q_{\rm tr}$ and refines the condition (\ref{condition}).
Furthermore, assuming $s$-channel momentum transfer, kinematics imposes 
$Q_{\rm tr}>2m_{\rm DM}$ so from Eq.~(\ref{4pi})
\be
\Lambda> \frac{m_{\rm DM}}{2\pi}\,,
\label{m2pi}
\ee
which is stronger than Eq.~(\ref{mover2pi}).
However,  the details of this condition depend case by case on the values
of $g_q, g_\chi$, and therefore on the details of the UV completion.

As an example of what the condition in Eq.~(\ref{condition}) means,
let us consider  the  hard scattering  process of the production
of two DM particles at  the parton-level
$q(p_1)+\bar q(p_2)\to \chi_1(p_3)+\chi_2(p_4)$.
Since the two $\chi$'s are produced on-shell, the energy injected into the diagram
needs to be $Q_{\rm tr}\geq 2m_{\rm DM}$, so $\Lambda\gtrsim Q_{\rm tr}$ is equivalent to
$\Lambda\gtrsim 2 m_{\rm DM}$,
which would be  much stronger than Eq.~(\ref{mover2pi}). However, the relevant process used for extracting lower limits on $\Lambda$ at the LHC is
when a single jet or photon is emitted from the initial state quark and this is the process we will consider in the next sections. We will proceed in two different ways. In the next section we will provide an estimate of the typical momentum transfer involved in the DM production associated to mono-jet or mono-photon, as a function of the transverse momentum and rapidity of the mono-jet or mono-photon. This calculation, even though not rigorous, will serve to provide an idea of the 
minimum value of $\Lambda$ compatible with the EFT approach. 
Subsequently, in Section \ref{sec:validity}, we will asses the validity of the EFT more precisely by studying the effect of the condition $Q_{\rm tr}<\Lambda$
on the cross sections for the production of DM plus mono-jet/photon, and by comparing
the cross section in the EFT and in the UV theory where the mediator has not been integrated out.

\section{An estimate of the momentum transfer}
\label{sec:estimate}

%
Many LHC searches for DM  are based on the idea of looking at events with missing energy plus a single jet or photon, emitted from the initial state. 
At the parton level  the  process is described by
\be
q(p_1)+\bar q(p_2)\to \chi_1(p_3)+\chi_2(p_4)+{\rm jet}(k) \,.
\ee
Let us see what happens if the energy exchanged is in the $s$-channel. Notice that, in the 
cases where the heavy mediator is exchanged in the $t$- or $u$-channels, one can always 
Fierz-rotate the corresponding non-renormalizable operator
to give rise to (a set of) operators where the momentum is transferred  in the $s$-channel.  
However, the basis of operators considered in experimental searches corresponds to
$s$-channel mediation only. 

In the $pp$ center of mass frame, the proton momenta take the
explicit form
$P_1=(\sqrt{s}/2,0,0,\sqrt{s}/2)$, $P_2=(\sqrt{s}/2,0,0,-\sqrt{s}/2)$,
where  $s$ is the center-of-mass energy.
 Ignoring the small transverse momenta of the partons,
 we can write the constituent quark momenta as fractions of these four-vectors
\be
p_1=x_1 P_1, \qquad p_2=x_2 P_2\,.
\label{momfractions}
\ee
 The four-momentum of the jet (assuming it is massless) is given in terms of transverse
momentum $p_{\rm T}$, pseudo-rapidity $\eta$ and azimuth angle $\phi$ by 
\be
k=(p_{\rm T}\cosh\eta, p_{\rm T}\cos\phi,p_{\rm T}\sin\phi,p_{\rm T}\sinh\eta)\, .
\ee
If the production of the DM takes place through the $s$-channel, then in the propagator it
will appear the quantity $Q_{\rm tr}^2-M^2$, where
\be
Q_{\rm tr}^2= (p_1+p_2-k)^2=
{x_1 x_2 s}-\sqrt{s}\,p_{\rm T}\left(x_1e^{-\eta}+x_2e^{\eta}\right)\,.
\label{Qtrasnfer}
\ee
The condition that $Q_{\rm tr}^2>0$ is equivalent to the condition
that the energy of the jet should be smaller than the energy of the parton it is emitted from.
The expression (\ref{Qtrasnfer}) is maximized at the rapidity value $e^{2\eta}=x_1/x_2$ corresponding to $
\left. Q_{\rm tr}^2\right\vert_{\rm max}
={x_1 x_2 s}-2\sqrt{s}\,p_{\rm T}\sqrt{x_1 x_2}$. 

\begin{figure}[t!]
\centering
\includegraphics[width=0.45\textwidth]{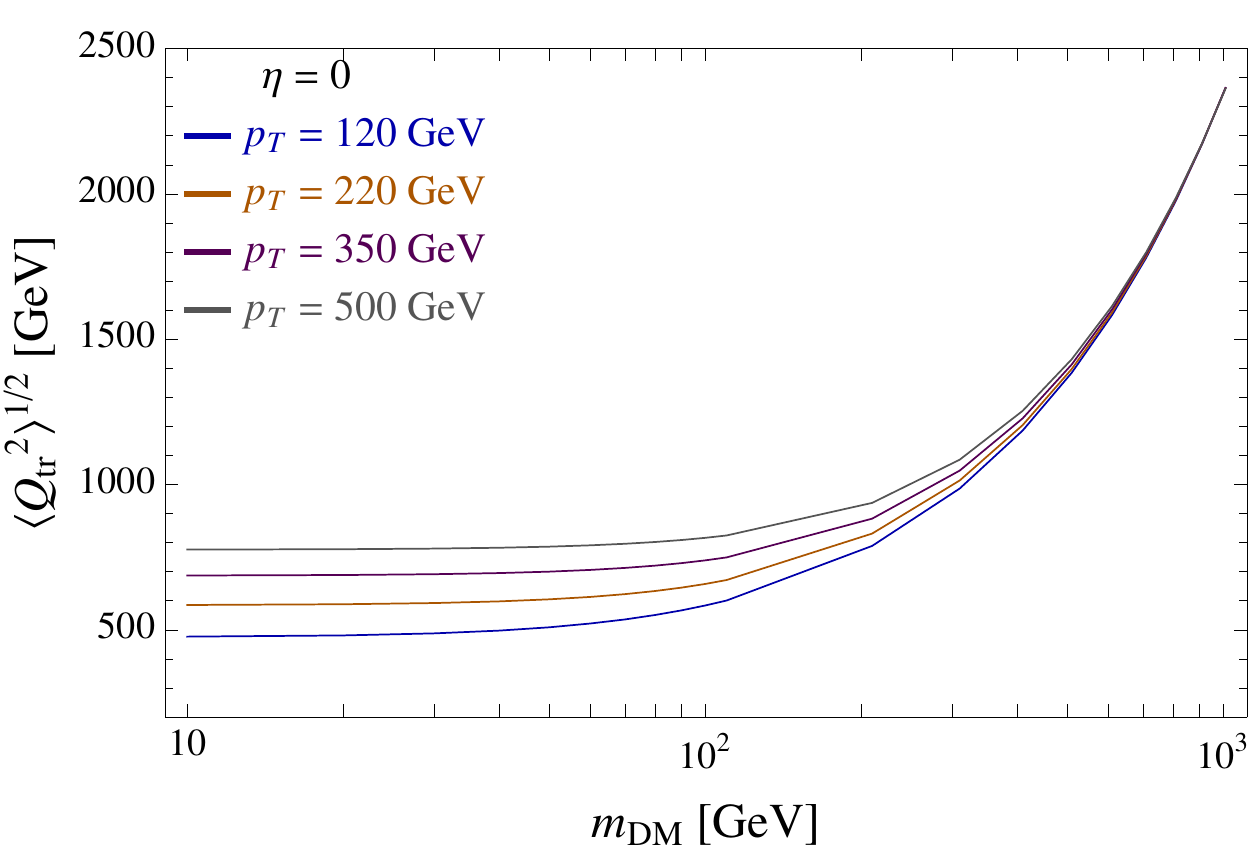}
\hspace{0.5cm}
\includegraphics[width=0.45\textwidth]{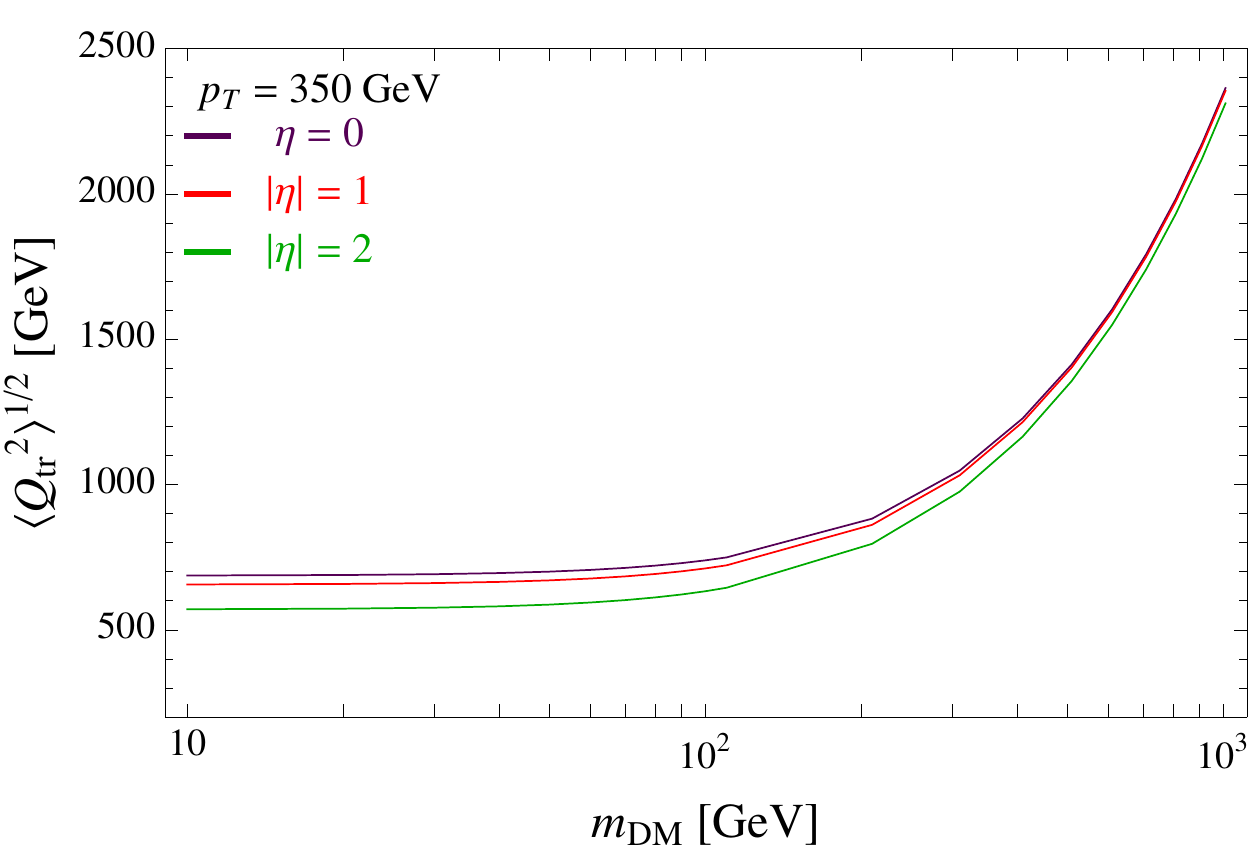}
\caption{\emph{{\small
The   momentum transfer in the $s$-channel in Eq.~(\ref{Qtrasnfer}),
weighted with PDFs, as a function of $m_{\rm DM}$, 
for different choices of $p_{\rm T}, \eta$ of the radiated jet.
 We considered $\sqrt{s}=8 \TeV$.
 }}} 
\label{fig:schannel}
\end{figure}

To assess the validity of the EFT, we first adopt  a procedure which,  albeit not rigorous, gives an
idea of the error one might make in adopting the EFT. The advantage of this procedure is that it is model-independent in the sense that it does not depend
on the particular UV completion of the EFT theory.
A simple inspection of the
expansion (\ref{ki}) tells us that the EFT is trustable only if $Q^2_{\rm tr}\ll M^2$ and we take for the typical value of  $Q_{\rm tr}$ the  square root of the averaged squared momentum transfer in the $s$-channel, where the average is computed  properly 
weighting with PDFs  \cite{pdf1}

\be
\label{avr}
\langle Q_{\rm tr}^2\rangle=\frac{\sum_q\int\de x_1\de x_2
\left[f_q(x_1)f_{\bar q}(x_2)+f_q(x_2)f_{\bar q}(x_1)\right]\theta(Q_{\rm tr}-2 m_{\rm DM})Q_{\rm tr}^2}
{\sum_q
\int\de x_1\de x_2
\left[f_q(x_1)f_{\bar q}(x_2)+f_q(x_2)f_{\bar q}(x_1)\right]\theta(Q_{\rm tr}-2 m_{\rm DM})}\,.
\ee
The integration in $x_1,x_2$ is performed Êover the kinematically allowed region $Q_{\rm tr}\geq 2m_{\rm DM}$
and we  have set the renormalization  and factorization scales  to $p_{\rm T}+2[m_{\rm DM}^2+p_{\rm T}^2/4]^{1/2}$, as often done by the LHC collaborations (see e.g.~Ref.~\cite{monojetATLAS2}). The results are plotted in Fig.~\ref{fig:schannel} as a function of the DM mass $m_{\rm DM}$ and 
for different choices of $p_{\rm T}$ and $\eta$ of the radiated jet. 
From Fig.~\ref{fig:schannel} we see that the lower the jet $p_{\rm T}$, the lower the momentum transfer is, and therefore  the better
the EFT will work.  
The same is true for smaller DM masses. These behaviors, which  are due to the fact we have  restricted the average of the mometum transfer to the
kinematically allowed domain, will be confirmed by a more rigorous approach in the next section. Notice that  $\langle Q_{\rm tr}^2\rangle^{1/2}$ is always larger than about 500 GeV, which poses a strong bound on the cutoff scale $\Lambda$: when  the coupling constants $g_q$ and $g_\chi$ are close to their perturbative regime, from the condition
(\ref{4pi}) we get $\Lambda \gtrsim 50$ GeV, but when the couplings are of order unity, one gets a much stronger bound $\Lambda \gtrsim 500$ GeV.

\section{Validity of the EFT approach}
\label{sec:validity}

The  tree-level differential cross sections for the hard scattering process 
$q\bar q\to \chi\chi$+gluon, in the UV (with interaction Lagrangian Eq.~(\ref{lagr})),
and in the EFT (with the operator Eq.~(\ref{os1})), are
\bea
\dfrac{\de^2\hat\sigma_{\rm eff}}{\de p_{\rm T}\de\eta}&=&
\frac{ \alpha_s}{36 \pi^2}
\frac{1}{p_{\rm T}}\frac{1}{\Lambda^4}
\frac{\left[Q_{\rm tr}^2-4m_{\rm DM}^2\right]^{3/2}\left[1+\frac{Q_{\rm tr}^4}{(x_1 x_2 s)^2}\right]}{\,Q_{\rm tr}}
\label{d2sigmaefflab}\,, \\
\dfrac{\de^2\hat\sigma_{\rm UV}}{ \de p_{\rm T}\de\eta}&=&
\frac{ \alpha_s}{36 \pi^2} \frac{1}{p_{\rm T}}\frac{ g_q^2g_\chi^2}{\left[Q_{\rm tr}^2-M^2\right]^2}
\frac{\left[Q_{\rm tr}^2-4m_{\rm DM}^2\right]^{3/2}
\left[1+\frac{Q_{\rm tr}^4}{(x_1 x_2 s)^2}\right]}{Q_{\rm tr}}\,,
\label{d2sigmaUVlab}
\eea
respectively, where $Q_{\rm tr}$ is given by Eq.~(\ref{Qtrasnfer}).
The corresponding cross sections initiated by the colliding protons are
\bea
\dfrac{\de^2\sigma_{\rm eff}}{\de p_{\rm T}\de\eta}&=&
\sum_q\int\de x_1\de x_2 
[f_q(x_1)f_{\bar q}(x_2)+f_q(x_2)f_{\bar q}(x_1)] 
\dfrac{\de^2\hat\sigma_{\rm eff}}{\de p_{\rm T}\de\eta}\,,\\
\dfrac{\de^2\sigma_{\rm UV}}{ \de p_{\rm T}\de\eta}&=&
\sum_{q}
\int\de x_1 \de x_2 
[f_q(x_1)f_{\bar q}(x_2)+f_q(x_2)f_{\bar q}(x_1)]
\dfrac{\de^2\hat\sigma_{\rm UV}}{ \de p_{\rm T}\de\eta}\,.
\eea
The explicit derivation of the Eqs.~(\ref{d2sigmaefflab})-(\ref{d2sigmaUVlab})  
can be found in  Appendix \ref{app:crosssect}. 
Throughout this work we willÊ identify the emitted gluon with the final jet observed experimentally. For  numerical results at NLO see Ref.~\cite{Fox:2012ru}.

The cross sections for the mono-jet processes are measured with a precision roughly of the order of 
10\%,
although this number can fluctuate due to many factors  (jet energy scale, PDFs, etc.).
However, as we are going to show, the effect of taking into account a cutoff scale can be larger than the precision of the cross section measurement, so the concern about the validity of the EFT approach is justified.

\subsection{The effect of the EFT cutoff}

Let us suppose we know  nothing about the UV completion of the EFT. Even so, we know that adopting only the lowest-dimensional operator of the EFT expansion is accurate only if the transfer energy is smaller than an energy scale of the order of  $\Lambda$, see Eqs.~(\ref{condition}), (\ref{ki}). However, up to what exact values of $Q_{\rm tr}/\Lambda$ is the EFT approach justified?  
Let us consider the ratio of  the cross section
obtained in the EFT by imposing the constraint  $Q_{\rm tr}<\Lambda$ on the
PDF integration domain, over the  cross section obtained in the EFT without such a constraint
\begin{figure}[t!]
\centering
\includegraphics[width=0.45\textwidth]{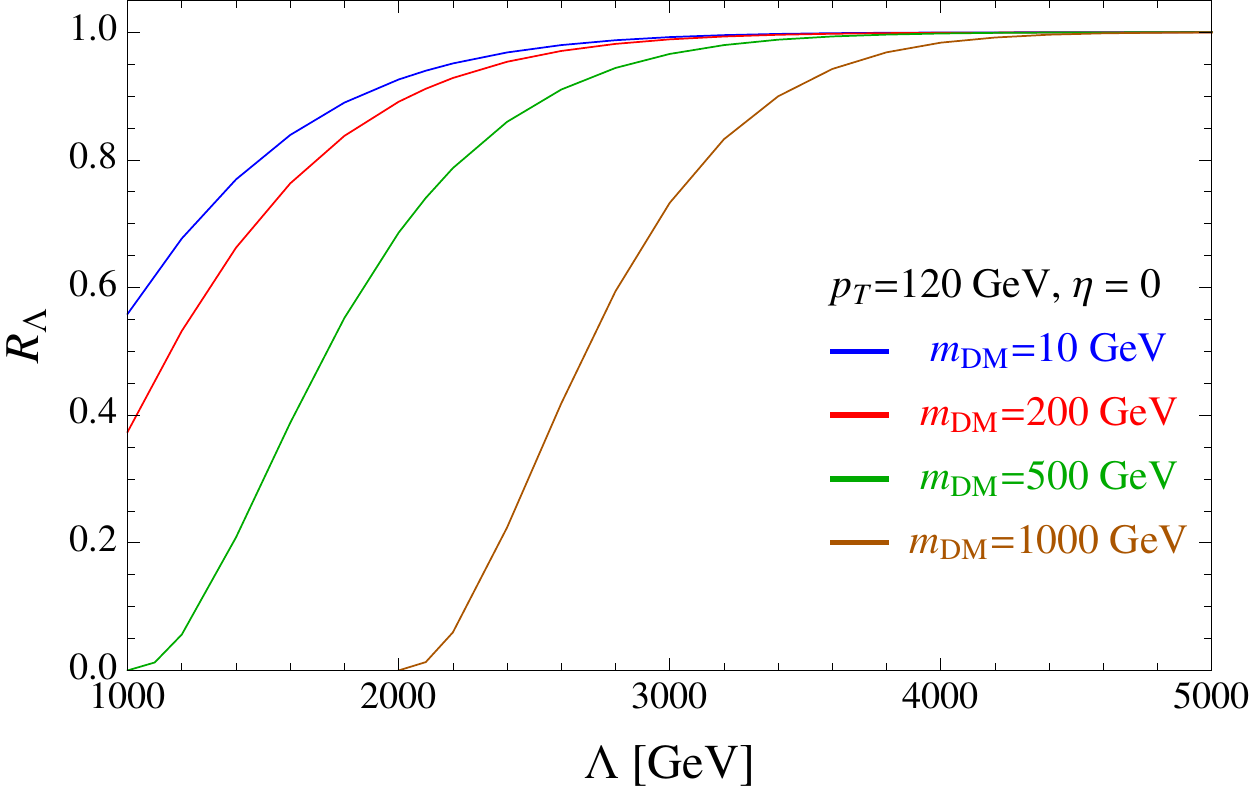}
\hspace{0.5cm}
\includegraphics[width=0.45\textwidth]{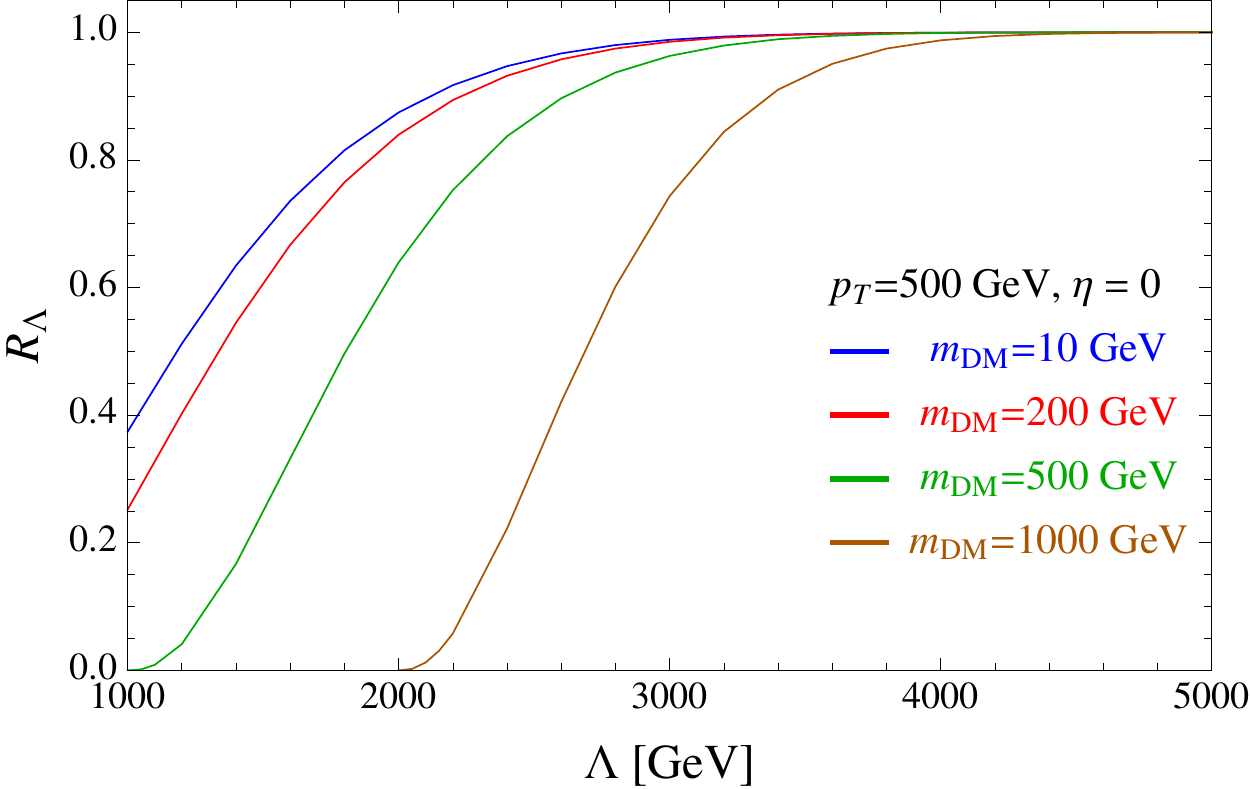}\\
\includegraphics[width=0.45\textwidth]{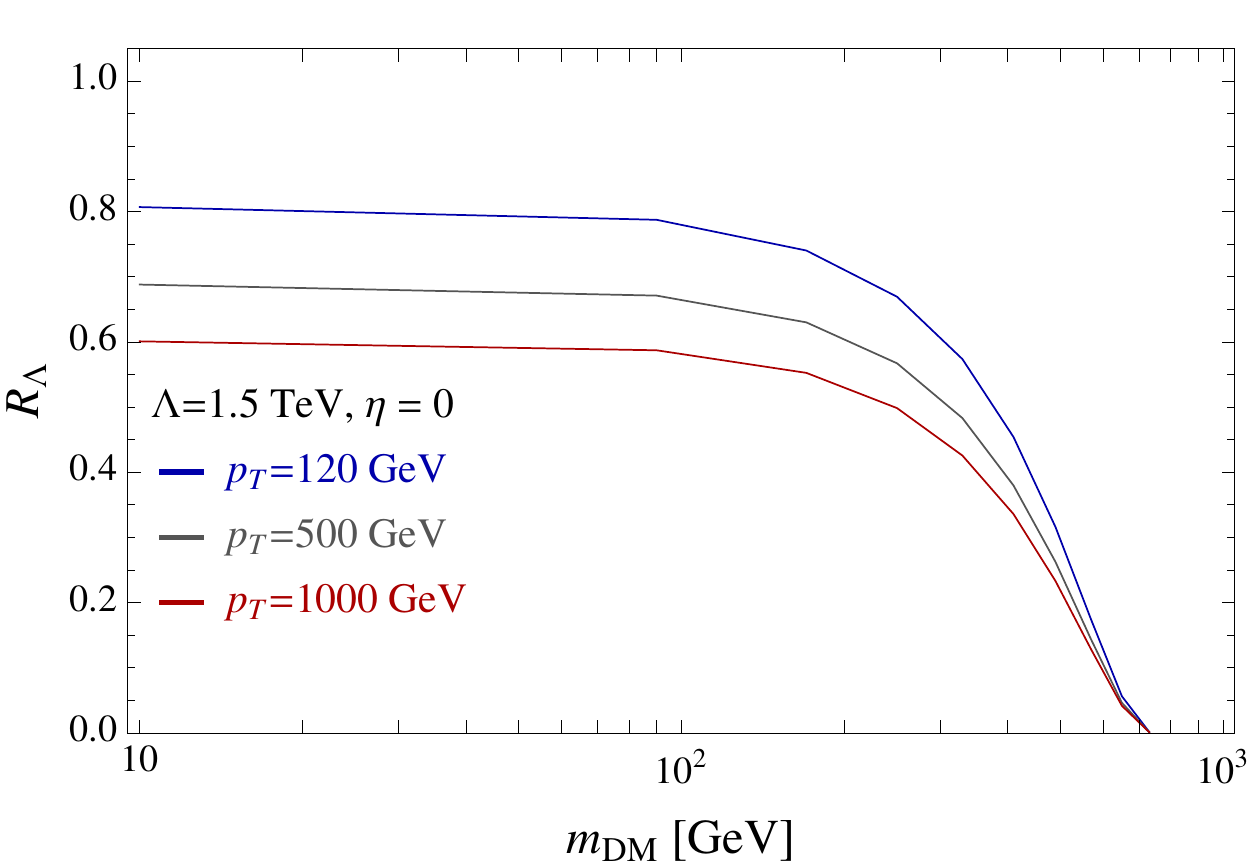}
\hspace{0.5cm}
\includegraphics[width=0.45\textwidth]{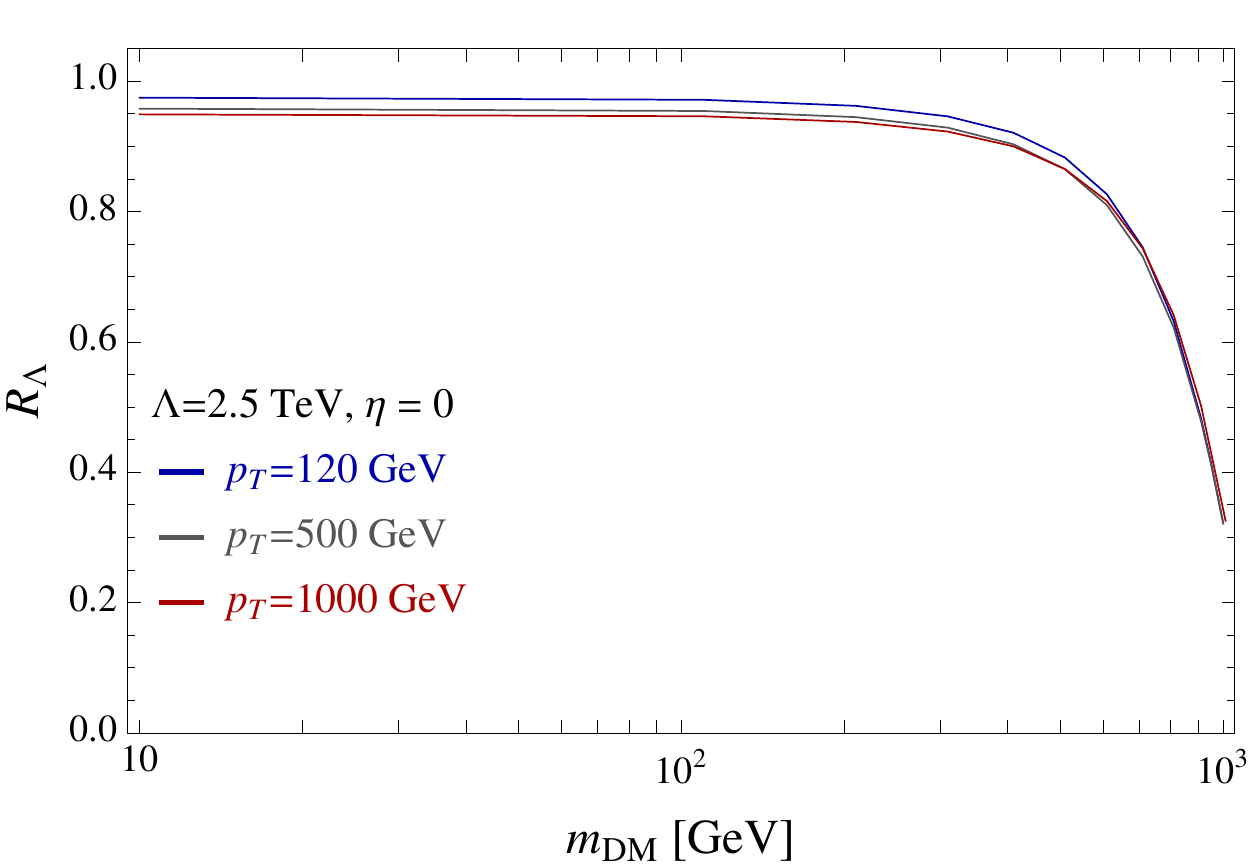}
\caption{ 
\emph{{\small 
The ratio $R_{\Lambda}$ defined in Eq.~(\ref{ratiolambda}) for $\sqrt{s}=8 \TeV, \eta=0$.
\emph{Top row:} 
$R_\Lambda$ as a function of
$\Lambda$, for various choices of $m_{\rm DM}$, for
 $p_{\rm T}=120 \GeV$ \emph{(left panel)}, $p_{\rm T}=500 \GeV$ \emph{(right panel}).
\emph{Bottom row:} 
 $R_{\Lambda}$ as a function of 
$m_{\rm DM}$, for various choices of $p_{\rm T}$, for
 $\Lambda= 1.5 \TeV$ \emph{(left panel)}, $\Lambda= 2.5 \TeV$ \emph{(right panel)}.
}}}
\label{fig:RLambda}
\end{figure}
\begin{figure}[t!]
\centering
\includegraphics[width=0.45\textwidth]{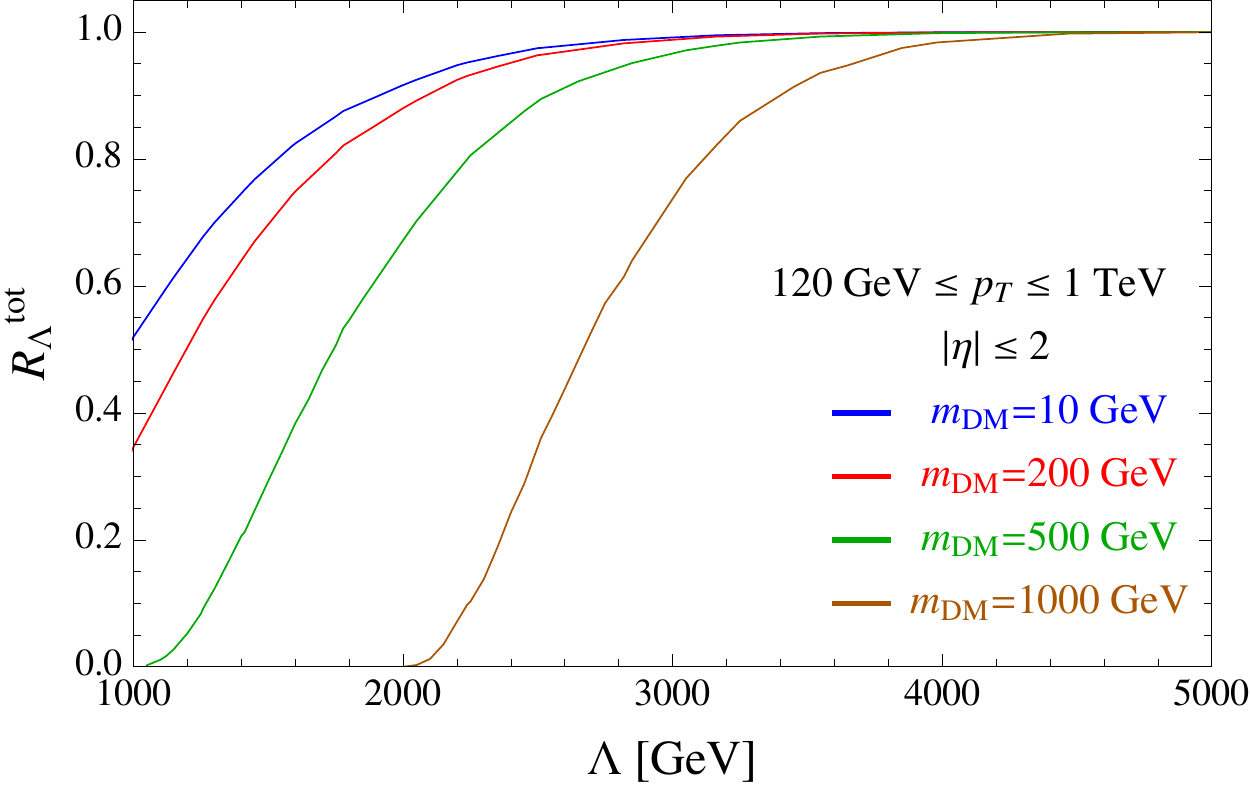}
\hspace{0.5cm}
\includegraphics[width=0.45\textwidth]{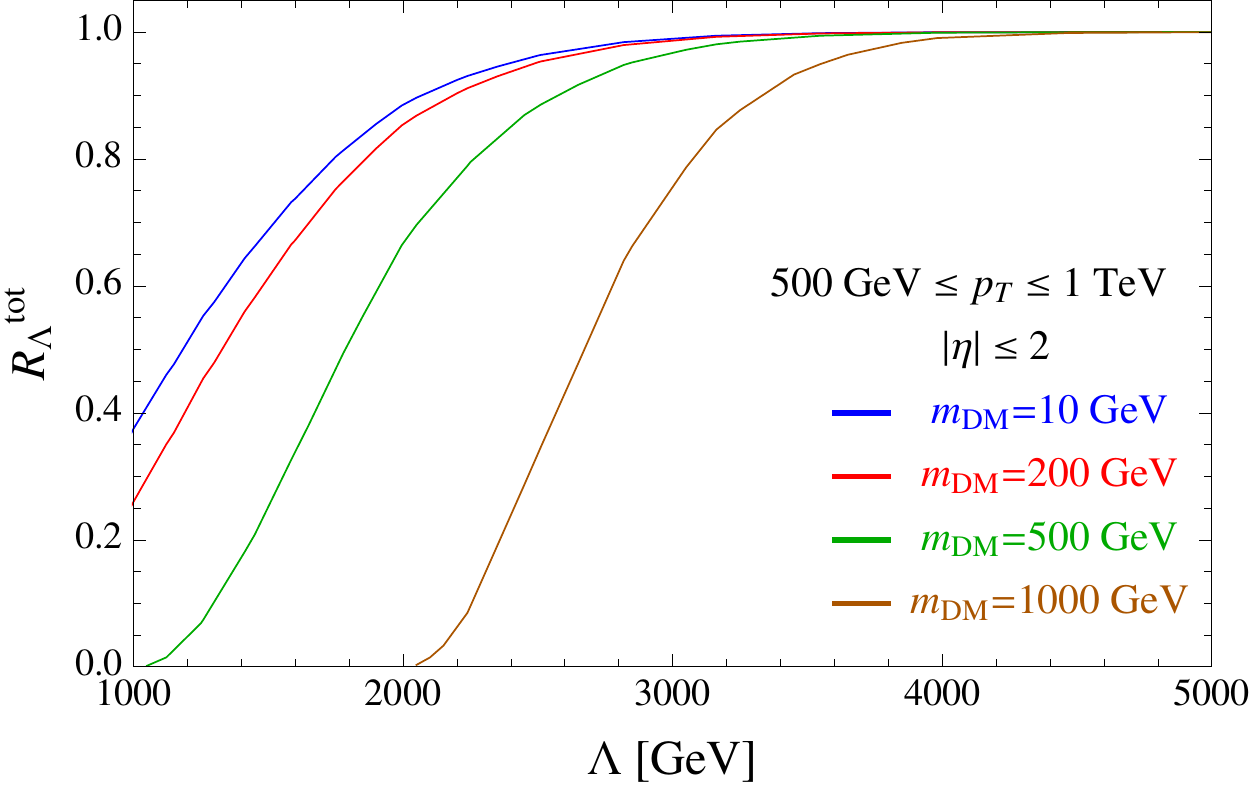}\\
\includegraphics[width=0.45\textwidth]{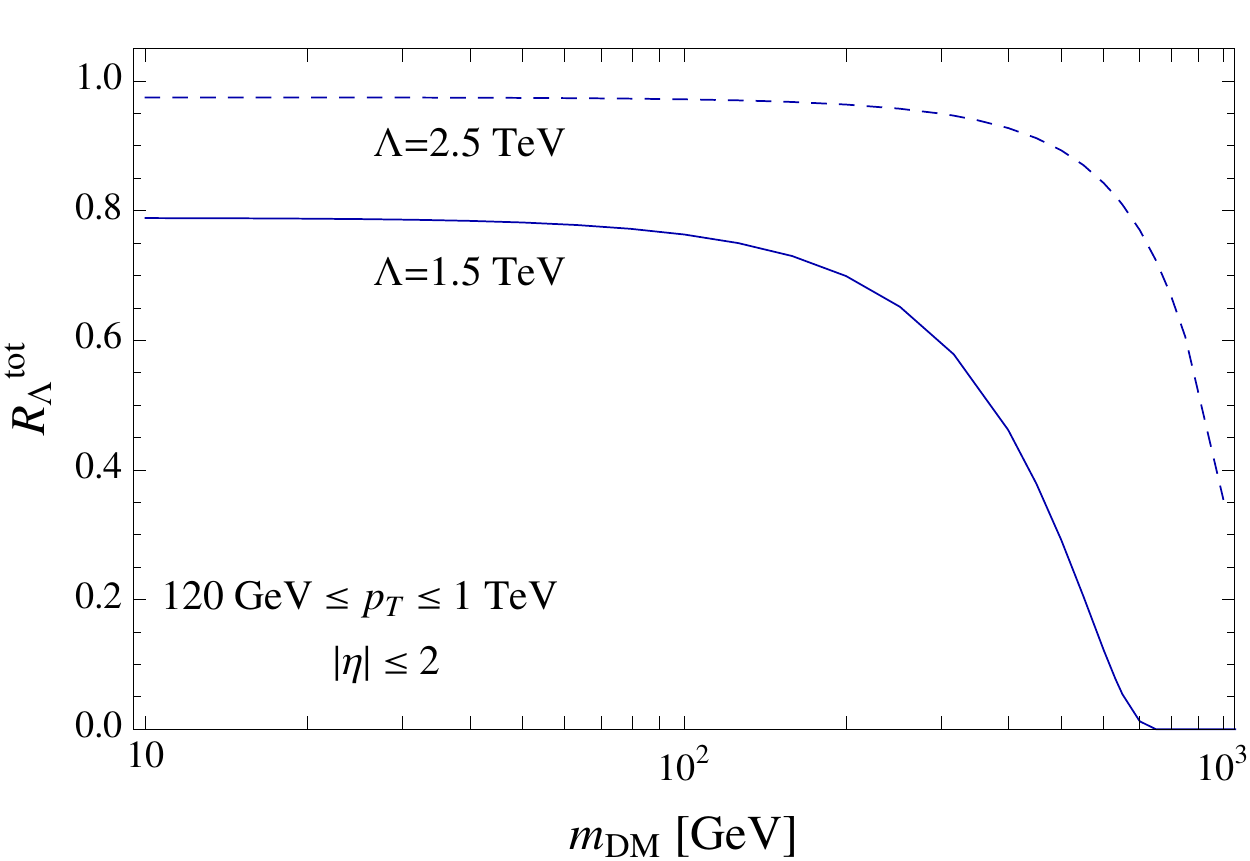}
\hspace{0.5cm}
\includegraphics[width=0.45\textwidth]{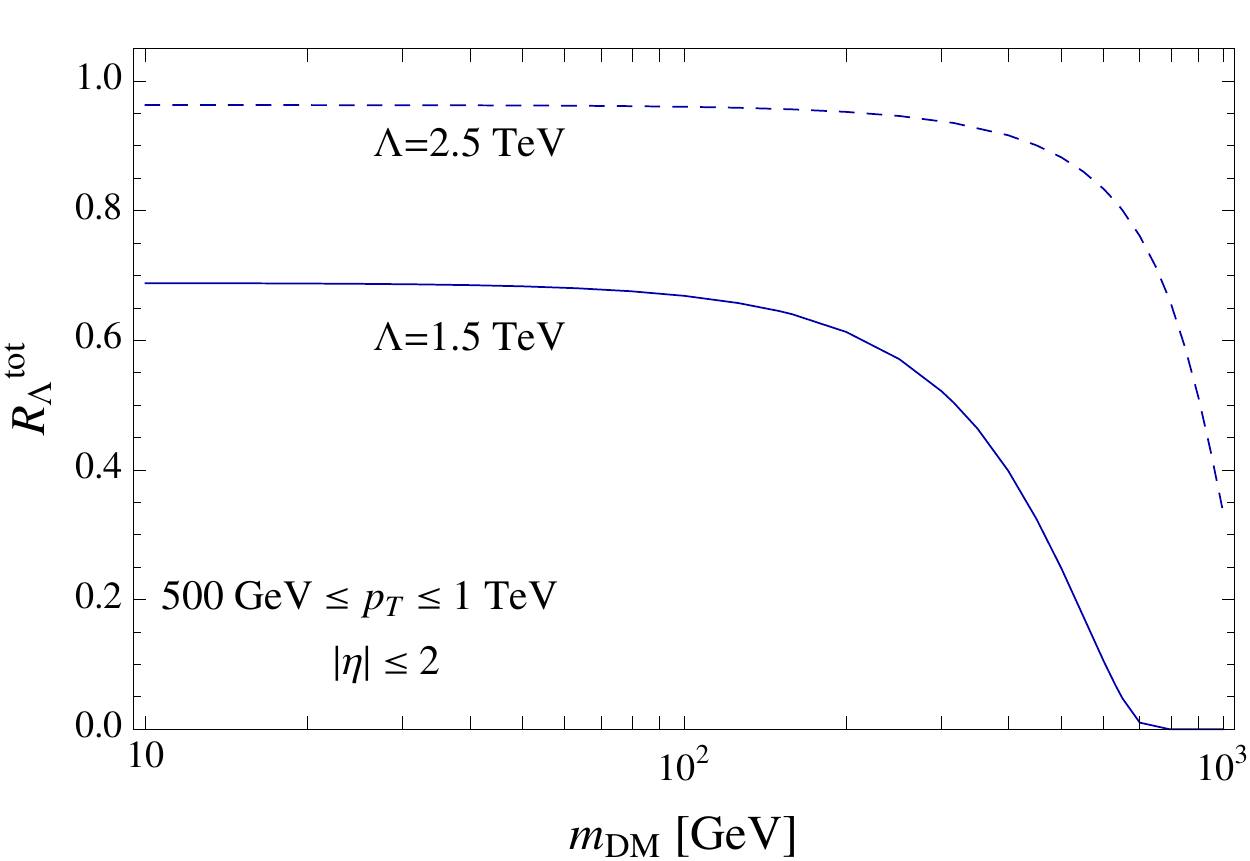}
\caption{ 
\emph{{\small
The ratio $R_{\Lambda}^{\rm tot}$ defined in Eq.~(\ref{ratiolambdatot}) for $\sqrt{s}=8 \TeV, |\eta|\leq 2$.
 \emph{Top row:} 
 $R_{\Lambda}^{\rm tot}$ as a function of 
$\Lambda$, for  $p_{\rm T}^{\rm min}=120 \GeV$  \emph{(left panel)},
$p_{\rm T}^{\rm min}=500 \GeV$  \emph{(right panel)}. 
\emph{Bottom row:} 
 $R_{\Lambda}^{\rm tot}$ as a function of 
$m_{\rm DM}$, for various choices of $\Lambda$, for   $ p_{\rm T}^{\rm min}=120 \GeV$  \emph{(left panel)},
$p_{\rm T}^{\rm min}=500 \GeV$  \emph{(right panel)}. 
}}}
\label{fig:RLambdatot}
\end{figure}
\be
R_{\Lambda}\equiv\frac{\left.\dfrac{\de^2\sigma_{\rm eff}}{\de p_{\rm T}\de\eta}\right\vert_{Q_{\rm tr}<\Lambda}}{\dfrac{\de^2\sigma_{\rm eff}}{ \de p_{\rm T}\de\eta}}\, .
\label{ratiolambda}
\ee
This ratio quantifies the fraction of the differential cross section for $q\bar q\to \chi\chi$+gluon,  for
given $p_{\rm T}, \eta$ of the radiated object,  mediated by the effective operator (\ref{os1}), where the momentum transfer
is below the scale $\Lambda$ of the operator.
Values of $R_{\Lambda}$ close to unity indicate that the effective 
cross section is describing
processes with sufficiently low momentum transfers, so the effective approach
is accurate.
On the other hand, a very small
$R_{\Lambda}$ signals that a significant error is made by extrapolating the effective description to a regime where it cannot be fully trusted, and where the neglected 
higher-dimensional operators  can give important contributions.

This ratio is plotted  in Fig.~\ref{fig:RLambda}  as a function of $\Lambda$ and $m_{\rm DM}$, for various choices  of  $p_{\rm T}$ and $\eta$.  
Our results indicate that if one would measure the cross section
for the mono-jet emission process within the EFT, but without taking into account that $Q_{\rm tr}$
should be bounded from above, one makes an error  which may even be very large,
depending on the values of the DM mass, the scale $\Lambda$ of the operator and the $p_{\rm T}, \eta$ of the emitted object.
Of course, the precise definition of the cutoff scale of an EFT is somewhat arbitrary,
with no knowledge of the underlying UV theory; therefore one
should consider the values of $R_\Lambda$ with a grain of salt.

To sum over the possible $p_{\rm T}, \eta$ of the jets, we integrate the cross sections
over values typically considered in the experimental searches and we can thus define
the following ratio of total cross sections
\be
R_\Lambda^{\rm tot}\equiv\frac{\sigma_{\rm eff}\vert_{Q_{\rm tr}<\Lambda}}{\sigma_{\rm eff}}
=\frac{\int_{p_{\rm T}^{\rm min}}^{1 \TeV}\de p_{\rm T}\int_{-2}^2\de \eta
\left.\dfrac{\de^2\sigma_{\rm eff}}{\de p_{\rm T}\de\eta}\right\vert_{Q_{\rm tr}<\Lambda}}
{\int_{p_{\rm T}^{\rm min}}^{1 \TeV}\de p_{\rm T}\int_{-2}^2\de \eta
\dfrac{\de^2\sigma_{\rm eff}}{\de p_{\rm T}\de\eta}}.
\label{ratiolambdatot}
\ee
As an example, we consider two cases:  $p_{\rm T}^{\rm min}=120 \GeV$, $500 \GeV$,
used in the signal regions SR1,  SR4 of \cite{monojetATLAS2}, respectively.
The results are shown in Fig.~\ref{fig:RLambdatot}. 
Notice  that both ratios $R_{\Lambda}, R_\Lambda^{\rm tot}$ get closer to unity for smaller DM masses, which confirms the qualitative analysis   on $\langle Q_{\rm tr}\rangle$ in Section
\ref{sec:estimate}, 
and also for larger $\Lambda$, when the effect of the cutoff becomes negligible.
On the other hand, $R_\Lambda$ goes to zero at $\Lambda=2m_{\rm DM}$, as
the phase space of DM pair production $Q_{\rm tr}\geq 2 m_{\rm DM}$ gets closed.
Notice also that the ratios involving  differential and total cross sections ($R_\Lambda$ and $R_\Lambda^{\rm tot}$) are very similar, as a consequence
of the fact that the integrands are very peaked at low $p_{\rm T}$ and at $\eta=0$.

We stress that this calculation does not rely on any specific UV completion of
the EFT, but it is completely rooted in the effective operator and the requirement of 
a consistent use of it within its range of validity.
Its only limitations are the lack of a precise identification of the cutoff scale
and that it applies to the case in which
the momentum transfer occurs in the $s$-channel.
The quantities in Eqs.~(\ref{ratiolambda})-(\ref{ratiolambdatot})
are not  directly deduced from actual data, but they require
 explicit analytical forms of the cross sections or MonteCarlo simulations of the events.
\begin{figure}[t!]
\centering
\includegraphics[width=0.45\textwidth]{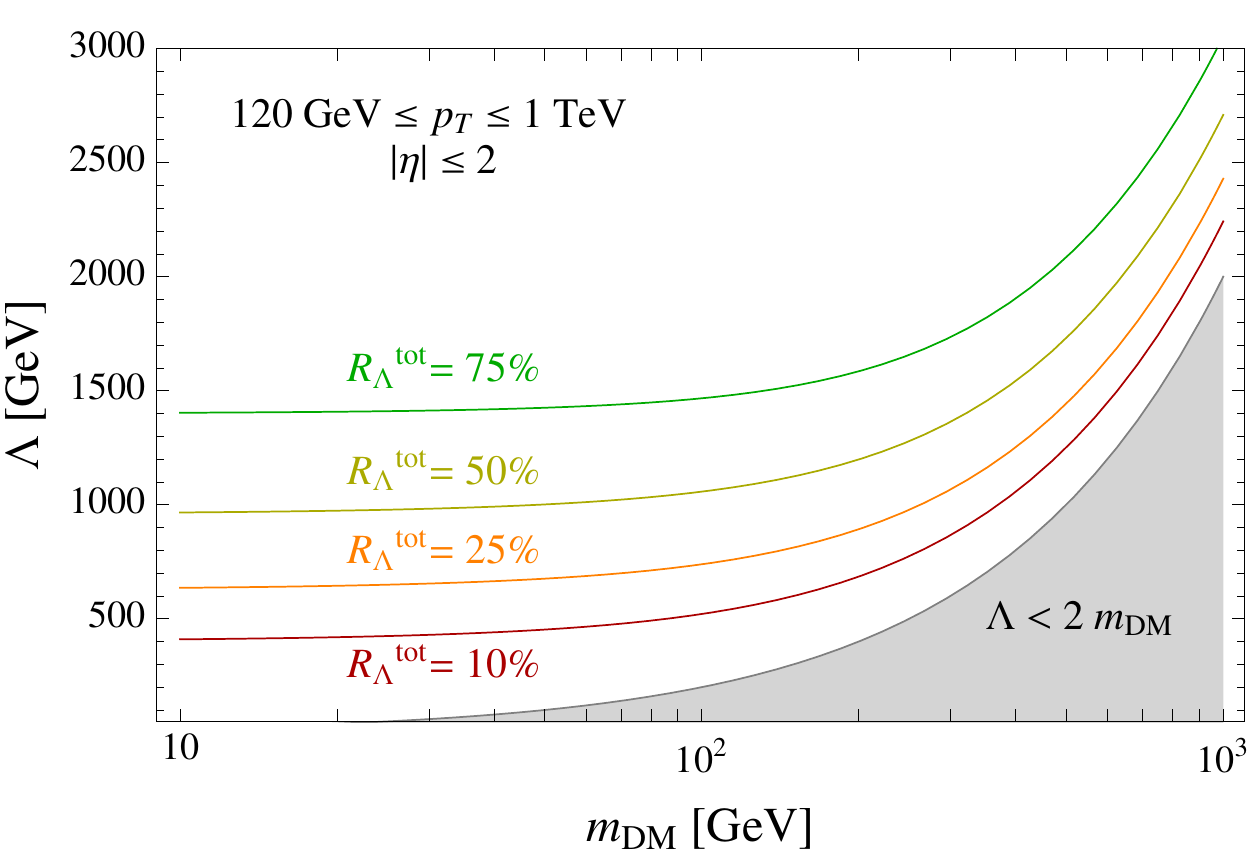}
\hspace{0.5cm}
\includegraphics[width=0.45\textwidth]{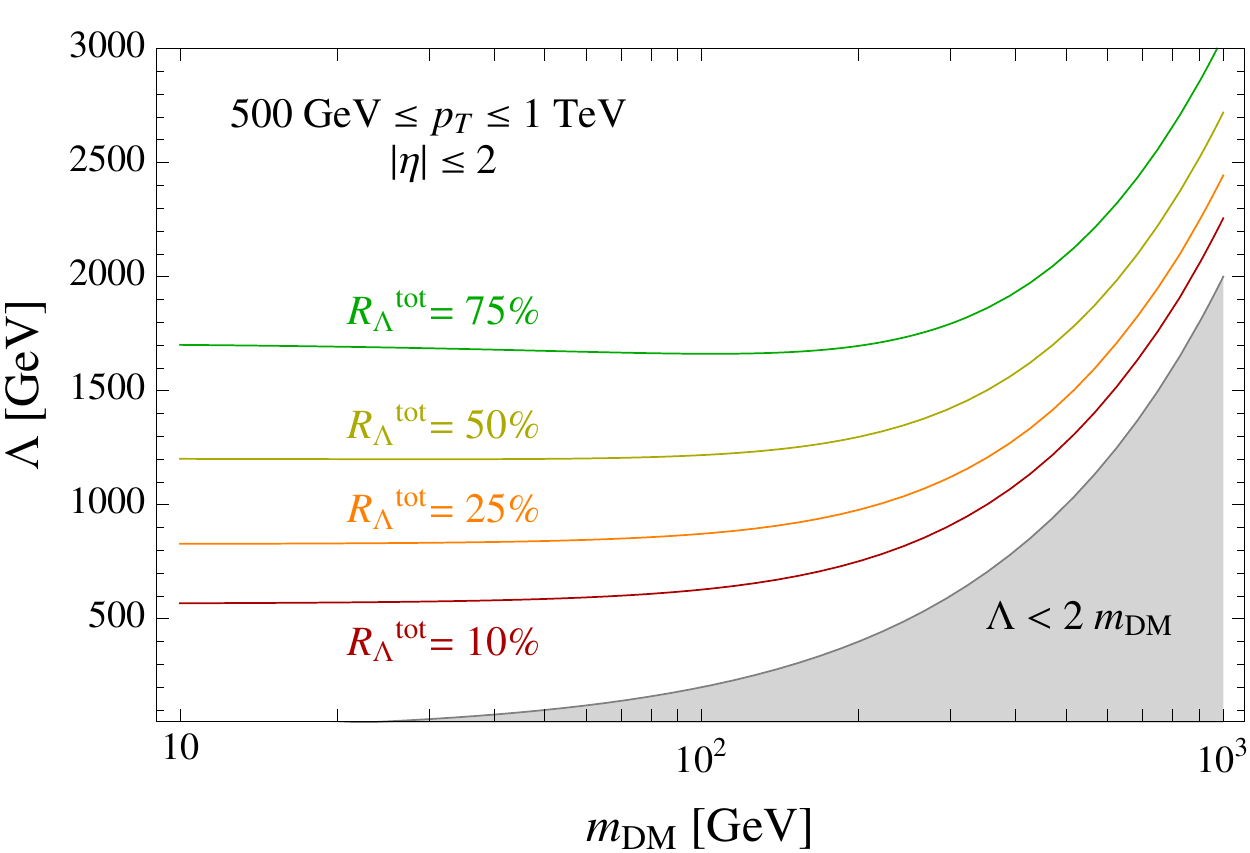}\\
\includegraphics[width=0.45\textwidth]{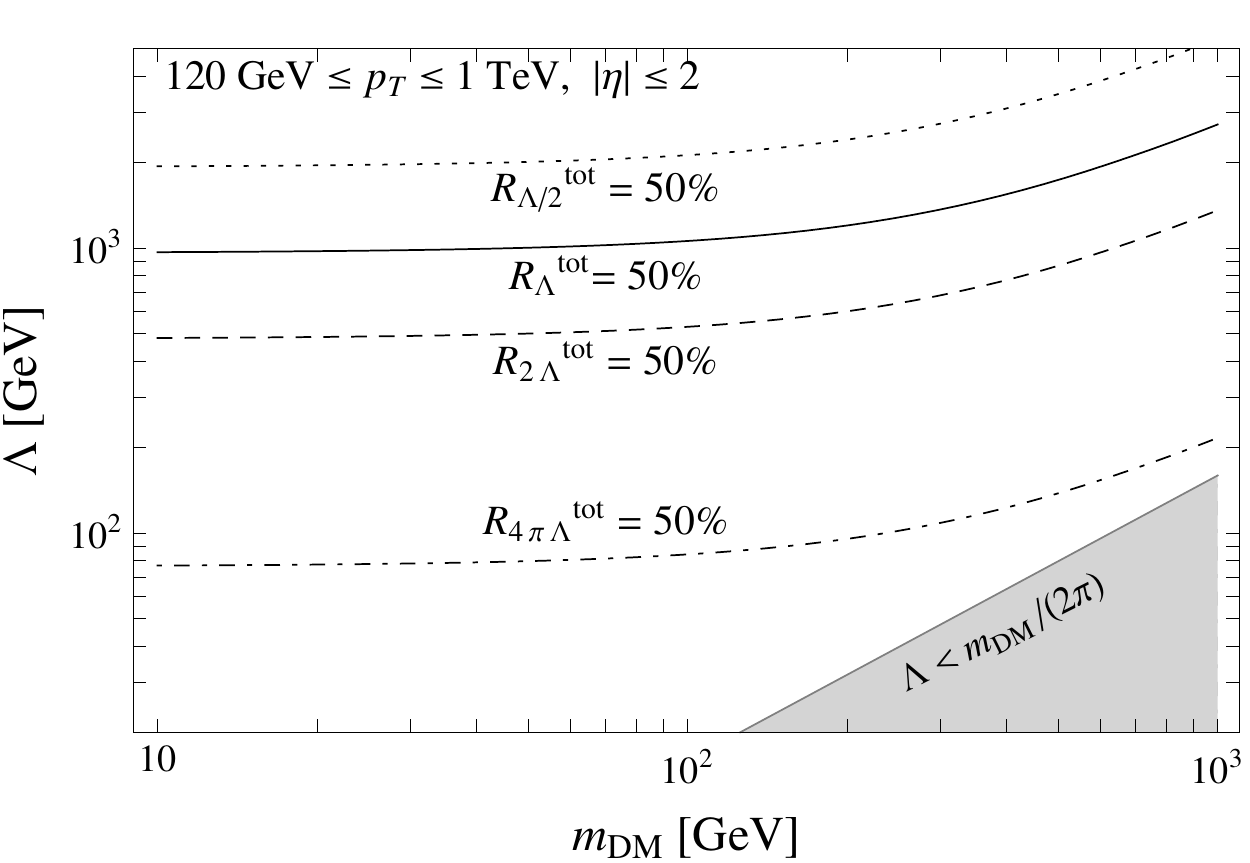}
\hspace{0.5cm}
\includegraphics[width=0.45\textwidth]{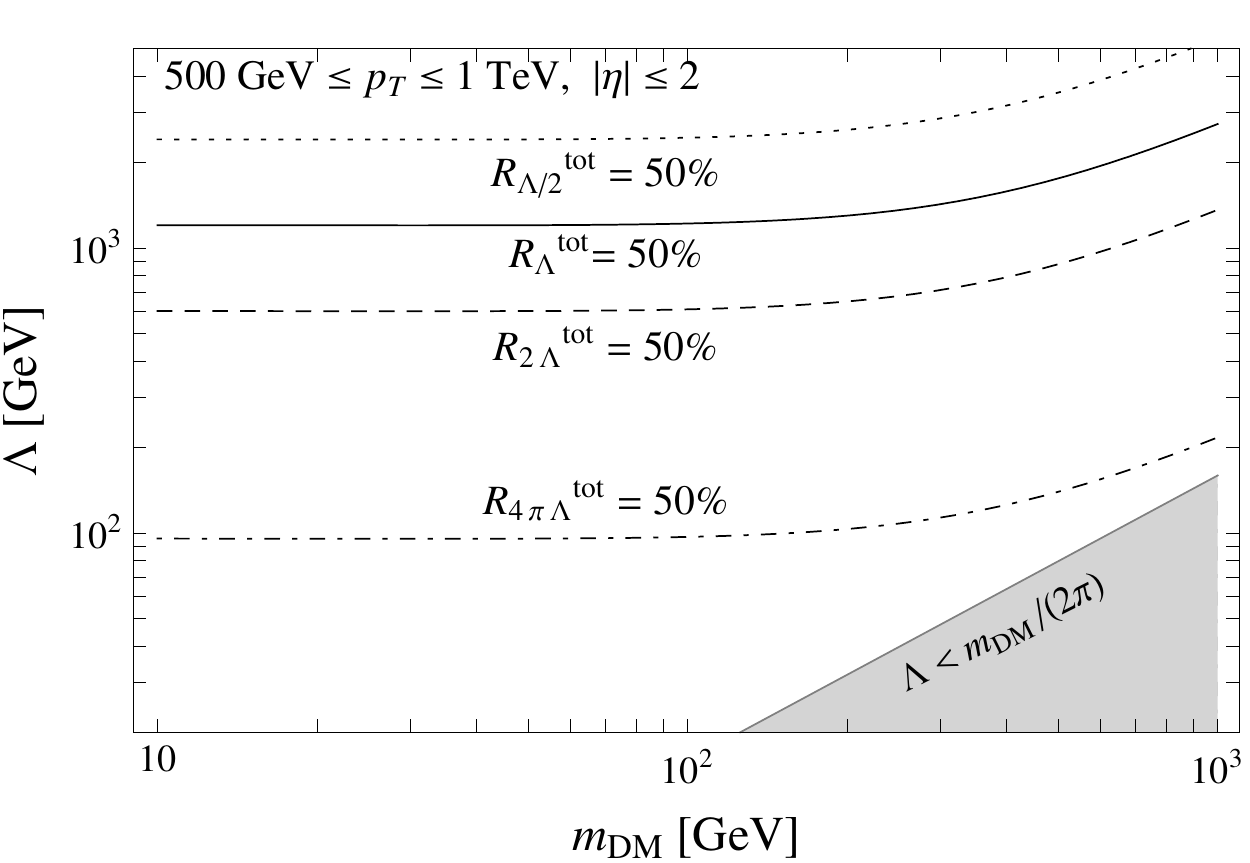}
\caption{ 
\emph{{\small
\emph{Top row:}
Contours for the ratio $R_{\Lambda}^{\rm tot}$, defined in Eq.~(\ref{ratiolambdatot}), on the plane
$(m_{\rm DM}, \Lambda)$. We set $\sqrt{s}=8 \TeV,
|\eta|\leq 2$ and $p_{\rm T}^{\rm min}=120 \GeV$ \emph{(left panel)}, 
$p_{\rm T}^{\rm min}=500 \GeV$ \emph{(right panel)}.
\emph{Bottom row:}
50\% contours for the ratio $R_{\Lambda}^{\rm tot}$, varying the cutoff $Q_{\rm tr}<\Lambda/2$
\emph{(dotted line)}, $\Lambda$ \emph{(solid line)}, 
$2\Lambda$ \emph{(dashed line)}, $4\pi\Lambda$ \emph{(dot-dashed line)}.
We have also shown the contour corresponding to $\Lambda<m_{\rm DM}/(2\pi)$
(see Eq.~(\ref{m2pi})),
which is often used as a benchmark for the validity of the EFT.
 We set $\sqrt{s}=8 \TeV,
|\eta|\leq 2$ and $p_{\rm T}^{\rm min}=120 \GeV$ \emph{(left panel)}, 
$p_{\rm T}^{\rm min}=500 \GeV$ \emph{(right panel)}.
}}}
\label{fig:RLambdacontours}
\end{figure}
We have also found  useful fitting functions for $R_\Lambda^{\rm tot}$ in the case 
$p_{\rm T}^{\rm min}=120$ GeV
\be
R_\Lambda^{\rm tot}= \left[1-e^{-1.273\left(\frac{\Lambda-2m_{\rm DM}}{1 \TeV}\right)^{1.752}}
\right]\left[1-e^{-1.326\left(\frac{\Lambda+2m_{\rm DM}}{1 \TeV}\right)^{0.903}}
\right]\,,
\label{fit1}
\ee
and in the $p_{\rm T}^{\rm min}=500$ GeV
\be
R_\Lambda^{\rm tot}= \left[1-e^{-1.265\left(\frac{\Lambda-2m_{\rm DM}}{1 \TeV}\right)^{1.820}}
\right]\left[1-e^{-0.714\left(\frac{\Lambda+2m_{\rm DM}}{1 \TeV}\right)^{1.385}}
\right]\,,
\label{fit2}
\ee
which are valid for $10 \GeV < m_{\rm DM} < 1 \TeV$, $800 \GeV<\Lambda<7 \TeV$, to better than
about 15\%.
The first factors in square brackets in Eqs.~(\ref{fit1})-(\ref{fit2}) are very mildly sensitive to the cut on  $p_{\rm T}$.
Of course, these results hold for the operator ${\cal O}_S$ in (\ref{os1}); for a different operator one would have a different fitting function.
The contours in the top row of  Fig.~\ref{fig:RLambdacontours}  indicate the regions in the parameter
space $(\Lambda, m_{\rm DM})$ where the description in terms of dim-6 effective operator
is accurate and reliable.
Even for very small DM masses, having $R_\Lambda^{\rm tot}$ at least 75\%, requires a cutoff scale at least above 1 TeV.

We reiterate that there is always some degree of arbitrariness  when defining precisely the
cutoff scale up to  which the EFT is reliable, as one does not know the details
of the UV physics integrated out.
This point reflects into the fact that the 
condition on the transfer momentum, 
see  Eq.~(\ref{4pi}), varies according to the values of $g_q, g_\chi$.
The effect  of varying the cutoff scale 
is shown in the bottom row of Fig.~\ref{fig:RLambdacontours},
for the representative contour $R_\Lambda^{\rm tot}=50\%$.
The extreme, and most conservative, situation $Q_{\rm tr}<4\pi\Lambda$, corresponding to couplings in the UV theory at the limit of the perturbative regime, is also shown.
Yet, the corresponding 50\% contour is above the limit
$\Lambda>m_{\rm DM}/(2\pi)$
(see Eq.~(\ref{m2pi})),
which is often used as a benchmark for the validity of the EFT.
This means that the parameter space regions of validity of the effective operator
approach can be smaller than commonly considered.

\subsection{Comparing the effective operator with a UV completion}

Let us now turn to  quantify the validity of the EFT by comparing cross sections for the production of DM plus mono-jet or mono-photon in the  simple example of a theory containing a DM particle $\chi$ and a heavy mediator
$S$ with the Lagrangian described in Eq. (\ref{lagr}) with its effective counterpart given by the operator in 
Eq.~(\ref{os1}).  The matching condition implies $\Lambda=M/\sqrt{g_q g_\chi}$.
Let us study the ratio of the cross sections obtained with the UV theory and with the effective operator
\be
r_{\rm UV/eff}\equiv\frac{\left.\dfrac{\de^2\sigma_{\rm UV}}{\de p_{\rm T}\de\eta}\right\vert_
{Q_{\rm tr}<M}}
{\left.\dfrac{\de^2\sigma_{\rm eff}}{\de p_{\rm T}\de\eta}\right\vert_{Q_{\rm tr}<\Lambda}}\, .
\label{ratioruveff}
\ee
This ratio quantifies the error of using the EFT, truncated at the lowest-dimensional operator,  with respect to its UV completion, for
given $p_{\rm T}, \eta$ of the radiated object.
Values of $r_{\rm UV/eff}$ close to unity indicate the effective operator is accurately 
describing the high-energy theory, whereas large values of $r_{\rm UV/eff}$
imply a poor effective description.

\begin{figure}[t!]
\centering
\includegraphics[width=0.45\textwidth]{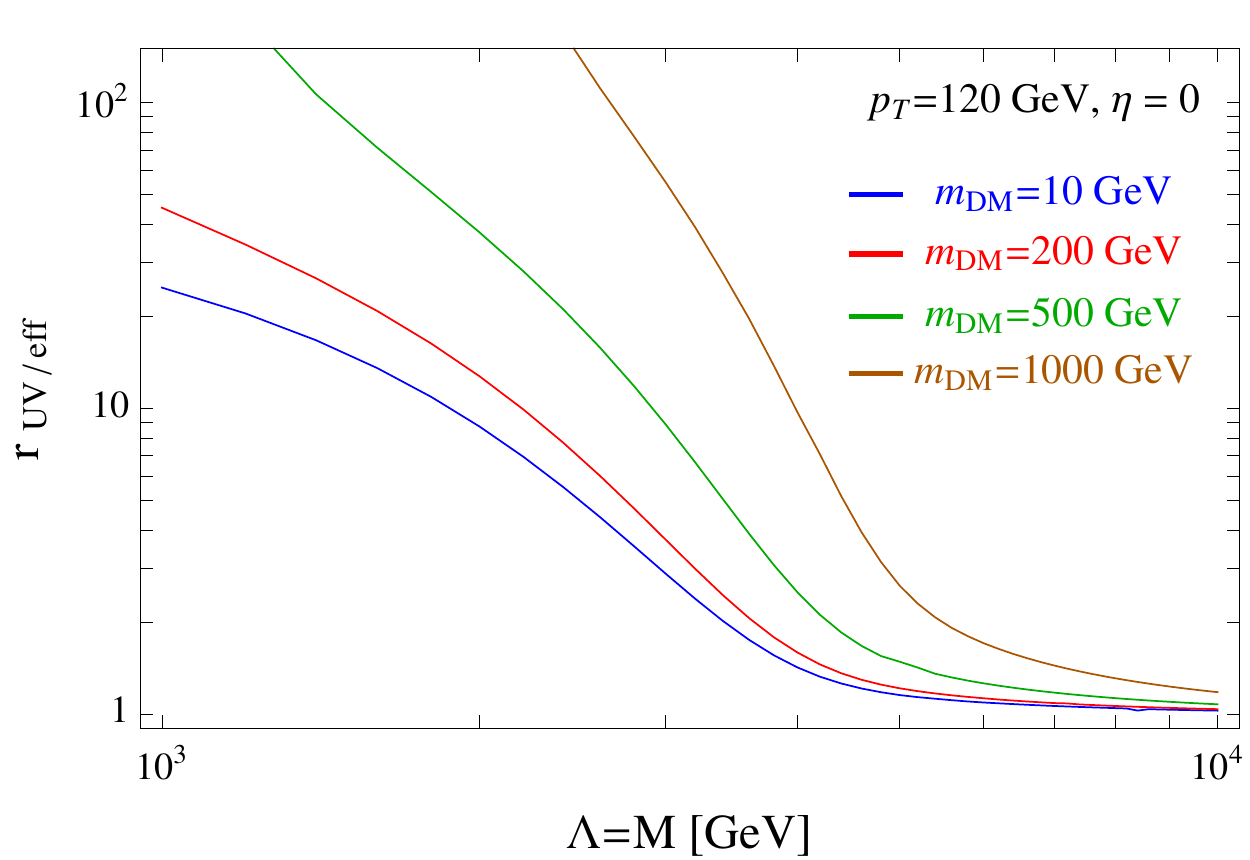}
\hspace{0.5cm}
\includegraphics[width=0.45\textwidth]{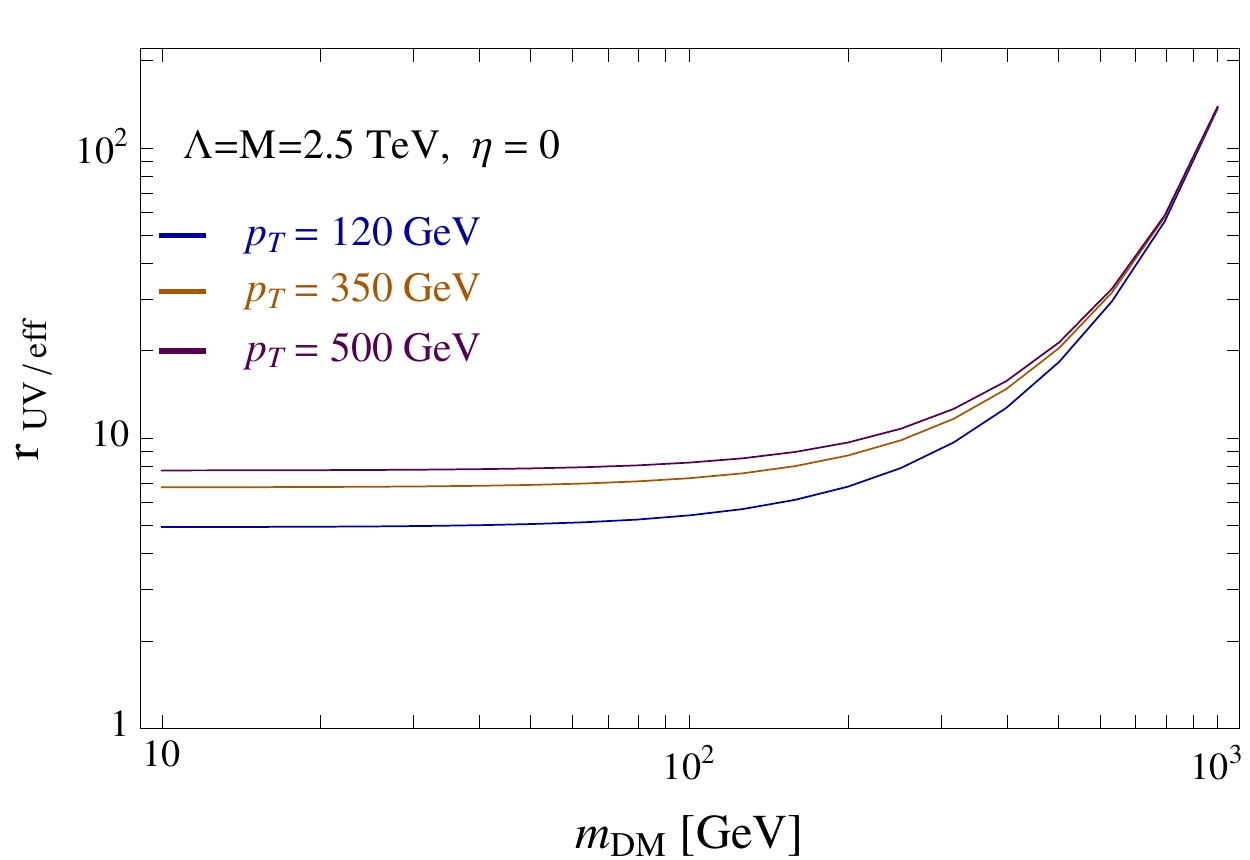}
\caption{\emph{{\small
The ratio $r_{\rm UV/eff}$ defined in Eq.~(\ref{ratioruveff}),
for $\sqrt{s}=8 \TeV, \eta=0$ and $M=\Lambda$ (corresponding to $g_q=g_\chi=1$).
\emph{Left panel:} $r_{\rm UV/eff}$ as a function of
 $\Lambda$  for  various choices
 of $m_{\rm DM}$ and $p_{\rm T}=120 \GeV$.
\emph{Right panel:} $r_{\rm UV/eff}$ as a function of
 $m_{\rm DM}$  for  various choices of $p_{\rm T}$, and $\Lambda=2.5 \TeV$.
}}} 
\label{fig:ratiod2sigma}
\end{figure}

For numerical integrations over the PDFs we have regularized the propagator introducing
a small width $\Gamma=(g_q^2+g_\chi^2)M/(8\pi)$ for the scalar mediator, which of course can
be larger in presence of additional decay channels.
The function $r_{\rm UV/eff}$ is plotted in Fig.~\ref{fig:ratiod2sigma}, for different 
choices of $p_{\rm T}, \eta, \Lambda, m_{\rm DM}$.
Once again, one can see  that the smaller  $p_{\rm T}$ and   $m_{\rm DM}$ are, the better the
 EFT works. 

\begin{figure}[t!]
\centering
\includegraphics[width=0.45\textwidth]{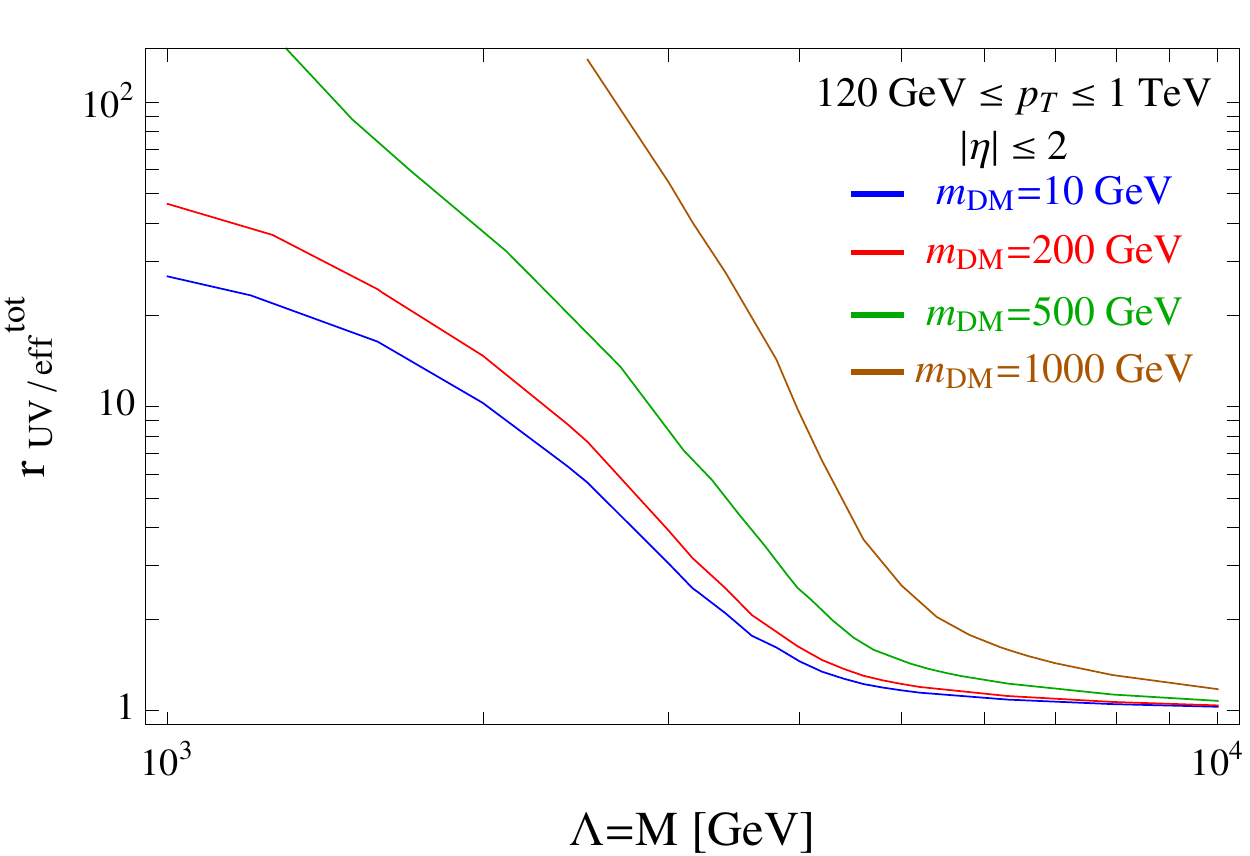}
\hspace{0.5cm}
\includegraphics[width=0.45\textwidth]{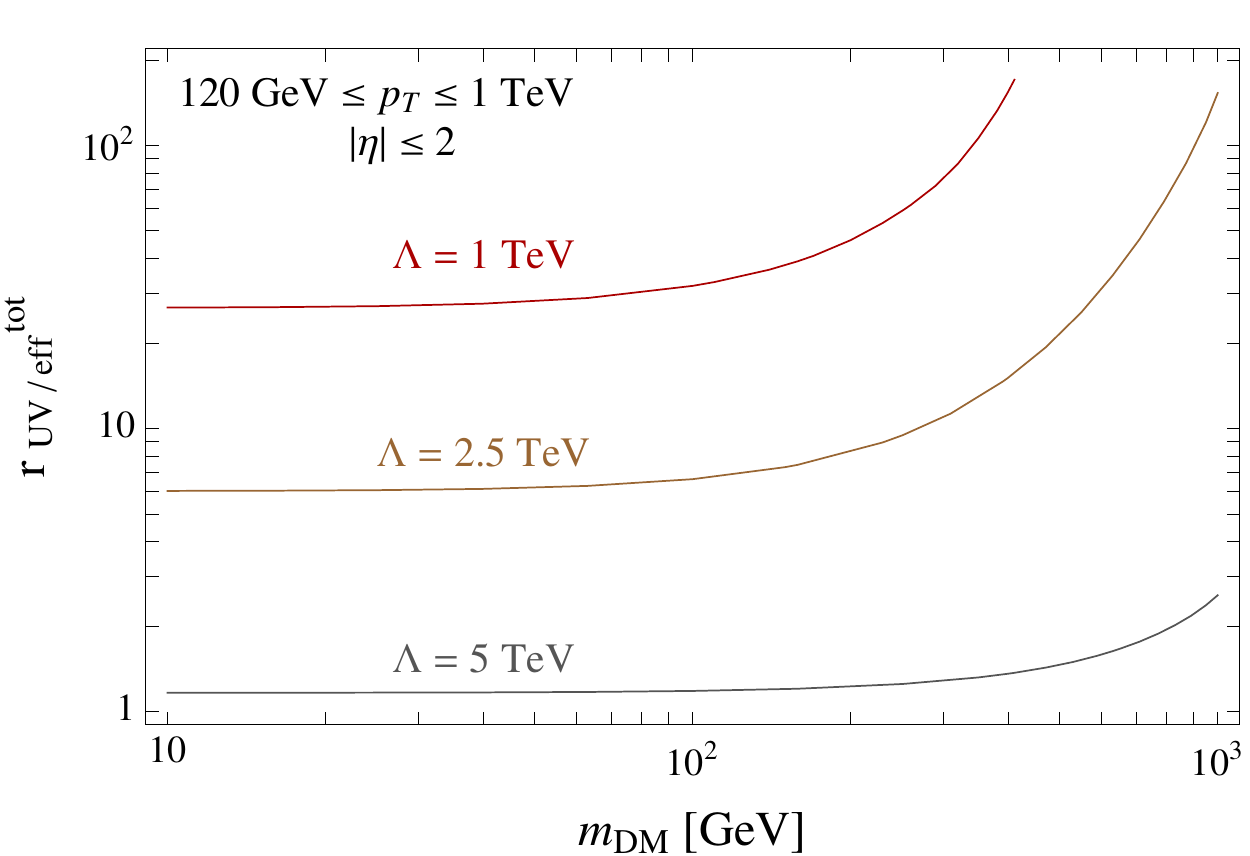}
\caption{\emph{{\small
The ratio $r_{\rm UV/eff}^{\rm tot}$ defined in Eq.~(\ref{ratioruvefftot}), as a function of
 $\Lambda$ \emph{(left panel)} and  $m_{\rm DM}$
\emph{(right panel)}. We have set $p_{\rm T}^{\rm min}=120 \GeV$, 
$|\eta|\leq 2$, $M=\Lambda$, $g_q=g_\chi=1$ and
 $\sqrt{s}=8 \TeV$. 
 }}} 
\label{fig:ratiosigma}
\end{figure}

Also, we can integrate over $p_{\rm T}, \eta$ using  cuts commonly used in the experimental
analysis (see e.g.~\cite{monojetATLAS2}): $p_{\rm T}\geq p_{\rm T}^{\rm min}, |\eta|\leq 2$

\be
r_{\rm UV/eff}^{\rm tot}\equiv\frac{\left.\sigma_{\rm UV}\right\vert_{Q_{\rm tr}<M}}{\left.\sigma_{\rm eff}\right\vert_{Q_{\rm tr}<\Lambda}}
=\frac{\int_{p_{\rm T}^{\rm min}}^{1 \TeV}\de p_{\rm T}\int_{-2}^2\de \eta
\left.\dfrac{\de^2\sigma_{\rm UV}}{\de p_{\rm T}\de\eta}\right\vert_{Q_{\rm tr}<M}}
{\int_{p_{\rm T}^{\rm min}}^{1 \TeV}\de p_{\rm T}\int_{-2}^2\de \eta
\left.\dfrac{\de^2\sigma_{\rm eff}}{\de p_{\rm T}\de\eta}\right\vert_{Q_{\rm tr}<\Lambda}}.
\label{ratioruvefftot}
\ee
The DM pair production is kinematically allowed for 
$Q_{\rm tr}>2 m_{\rm DM}$; furthermore, when dealing with the processes with
mediator exchange, 
one also has to require $ Q_{\rm tr} <M$  to avoid on-shell mediator. Therefore we have   worked with the condition $
2 m_{\rm DM}< Q_{\rm tr} <M$, 
 which can only be satisfied if $m_{\rm DM}<M/2$.
The results for $r_{\rm UV/eff}^{\rm tot}$  are plotted in Fig.~\ref{fig:ratiosigma}. Again, we see that one needs a cutoff scale $\Lambda$
at least larger than about a few TeV in order for the ratios  $r_{\rm UV/eff}$ and $r_{\rm UV/eff}^{\rm tot}$  to be of order unity, the best case being attained for the lowest DM masses. 
As in the previous subsection, the ratios involving  differential and total cross sections ($r_{\rm UV/eff}$ and $r_{\rm UV/eff}^{\rm tot}$) are very similar, as a consequence
of the fact that the integrands are very peaked at low $p_{\rm T}$ and at $\eta=0$.

As a final remark of this section, notice  that $\sigma_{\rm UV}$ turns out to be easily much bigger than $\sigma_{\rm eff}$. This means that interpreting the mono-jet data in terms of effective
operator or in terms of mediator exchange can make a big difference.
In particular, it implies that there can be placed  more stringent bounds on the mediator mass of the simple model than  on the cutoff scale of the effective operators. We also expect
the direct exclusion bounds from the negative searches of heavy mediators (e.g. di-jet searches) 
to play an important role.

\section{Conclusions}
\label{sec:conclusions}
The EFT approach is commonly used to study the indirect signatures of the production 
of DM particles at LHC. While this approach has the undeniable advantage of being independent of the plethora of models of DM, its validity  has to be scrupulously analyzed as the momentum transfers involved in the mono-jet and mono-photon searches can be rather large. 

In this paper we have introduced various quantities which can help in assessing the validity of the EFT approach for DM searches. Some of these quantities have the virtue of being independent of the UV completion of the DM model. For instance, for a specific operator connecting DM particles with quarks,
we have introduced the ratio $R_\Lambda^{\rm tot}$ (see Eq.~(\ref{ratiolambdatot})), which is a measure of the error one would make by extrapolating the
effective description to a regime with very high momentum transfers, 
where it cannot be fully trusted.
It does not rely on any specific UV completion of
the EFT, but simply follows from the requirement of 
using the effective approach consistently, for momentum transfers below the cutoff scale
of the operator. 
Given its  large range of applicability, we have  provided simple fitting functions Eqs.~(\ref{fit1})-(\ref{fit2}) which may be used to set general criterions on the validity of the EFT approach. 

We have also studied in detail a specific example of UV physics giving rise
to a given operator at low energies. In this model, the SM sector and the DM particles
are connected through a heavy scalar mediator. The 
ratio $r_{\rm UV/eff}^{\rm tot}$ (see Eq.~(\ref{ratioruvefftot}))
relates the cross section
for mono-jet production obtained with the effective operator and with its possible UV completion.
The analysis 
confirms the general conclusions one can draw from the model-independent
quantities, that is: the validity of the EFT approach in studying indirect signals of DM at the LHC requires the cutoff scale $\Lambda$ to be
larger than about 1 TeV (unless the couplings constants involved in the processes are close to the non-perturbative regime).
Increasing the mass of the DM increases the lower bounds on $\Lambda$. The reason for this behavior is clear: larger DM masses imply larger momentum transfers which, in turn, require larger values of the EFT cutoff scale $\Lambda$.

Thus, we conclude that the use of EFT for DM searches in a highly energetic environment, such as  the LHC, should be handled with care. This point has been already remarked in various references
\cite{Goodman:2010ku, Dreiner:2012xm, Cotta:2012nj,  Shoemaker:2011vi, Fox:2012ru, Fox:2011fx}, 
and our results are in general agreement with theirs.
The results of the experimental searches carried out in the EFT language need to be confronted, 
consistently  and case by case, with the validity of the effective description itself.
Although we have restricted our analysis to a specific operator, it is clear that the very same logic
discussed in this paper is equally applicable to any other situation or operator.
In particular, it would be interesting also to investigate the case of scalar DM particles,
which has not yet been studied by the experimental LHC collaborations.

\section*{Acknowledgments}
We thank Johanna Gramling, Giuseppe Iacobucci, Xin Wu for many useful interactions on the subject of dark matter searches at ATLAS, and 
Giovanni Ridolfi and Alessandro Vichi for interesting conversations.
ADS acknowledges partial support from the  European Union FP7  ITN INVISIBLES (Marie Curie Actions, PITN-GA-2011-289442).

\appendix
\section{Three-body Cross Sections}
\label{app:crosssect}

In this Appendix we show the details of the calculations of the tree-level cross sections for
 the hard scattering process $q(p_1)+\bar q(p_2)\to \chi(p_3)+\chi(p_4)+\gamma/g(k)$, where the photon/gluon is emitted from the initial state quark (of charge $Q_qe$). 
 For simplicity we show explicitly the calculation for the photon emission, the case of gluon
emission amounts to a simple rescaling of the overall coefficient (see  the end
of the Appendix).

The differential
 cross section    is generically given by
\begin{equation}
d\hat\sigma=\frac{\sum\overline{|{\cal M}|^2}}{4(p_1\cdot p_2)}
\de \Phi_3\,,
\end{equation}
where  the three-body phase space is
\be
\de \Phi_3=
(2\pi)^4\delta^{(4)}(p_1+p_2-p_3-p_4-k)\frac{{\rm d}\textbf{p}_3}{(2\pi)^32p_3^0}
\frac{{\rm d} \textbf{p}_4}{(2\pi)^32p_4^0}\frac{{\rm d}\textbf{k}}{(2\pi)^32k^0}\,.
\ee
The amplitudes for the processes with the exchange of a scalar of mass $M$, whose interactions
are described by the Lagrangian in Eq.~(\ref{lagr}), and with the effective operator ${\cal O}_S$
 in Eq.~(\ref{os1})  are
\bea
{\cal M}_{\rm UV}&=&
\frac{i Q_q e g_q g_\chi}{(p_1+p_2-k)^2-M^2}
\left[\bar v(p_2)\left(
\frac{2 p_2\cdot \epsilon^* +\slashed{k}\slashed{\epsilon}^*}{2 p_2\cdot k}
-\frac{2 p_1\cdot \epsilon^* +\slashed{\epsilon}^*\slashed{k}}{2 p_1\cdot k}
\right) u(p_1)\right][\bar u(p_3)v(p_4)]\,,\nn\\
&&\\
{\cal M}_{\rm eff}&=&
-\frac{i Q_q e}{\Lambda^2}
\left[\bar v(p_2)\left(
\frac{2 p_2\cdot \epsilon^* +\slashed{k}\slashed{\epsilon}^*}{2 p_2\cdot k}
-\frac{2 p_1\cdot \epsilon^* +\slashed{\epsilon}^*\slashed{k}}{2 p_1\cdot k}
\right) u(p_1)\right][\bar u(p_3)v(p_4)]\,,
\eea
respectively.
The squared amplitudes, averaged over initial states (spin and color) and summed over final states, are
\bea
\sum\overline{|{\cal M}_{\rm UV}|^2}&=&
\frac{4}{3}Q_q^2e^2 g_q^2 g_\chi^2
\frac{[(p_3\cdot p_4)-m_{\rm DM}^2]\left[(k\cdot(p_1+p_2))^2-2(p_1\cdot p_2)(k\cdot p_1+k\cdot p_2-p_1\cdot p_2) \right]}
{[(p_1+p_2-k)^2-M^2]^2(k \cdot p_1)(k\cdot p_2)}\,,
\label{MsquaredUV}\nn\\
&&\\
\sum\overline{|{\cal M}_{\rm eff}|^2}&=&
\frac{4}{3}\frac{ Q_q^2e^2}{\Lambda^4}
\frac{[(p_3\cdot p_4)-m_{\rm DM}^2]\left[(k\cdot(p_1+p_2))^2-2(p_1\cdot p_2)(k\cdot p_1+k\cdot p_2-p_1\cdot p_2)\right]}
{(k \cdot p_1)(k\cdot p_2)}\,.\nonumber\\
&&
\label{Msquaredeff}
\eea
At this stage one can proceed in two equivalent ways: either by evaluating the scalar products
and the phase space directly in the lab frame, or by computing them in the center-of-mass frame
and then boosting the result to the lab frame. We show the details for the latter procedure,
but we have also carried out the calculation in the former way and checked the agreement.

Let us first write down the four-momenta in components in the center-of-mass (c.o.m.) frame of the
two colliding partons, carrying equal momentum fractions $x_1=x_2\equiv x$ of the incoming
protons
\bea
p_1&=&x \frac{\sqrt{s}}{2}(1,0,0,1)\,,\qquad 
p_2=x \frac{\sqrt{s}}{2}(1,0,0,-1)\,, \qquad
k= x\frac{\sqrt{s}}{2}(z_0,z_0\hat k)\, ,\\
p_3&=&x \frac{\sqrt{s}}{2}(1-y_0,\sqrt{(1-y_0)^2-a^2}\hat p_3 )\,,\quad
p_4=x \frac{\sqrt{s}}{2}(1+y_0-z_0,\sqrt{(1+y_0-z_0)^2-a^2}\hat p_4)\, ,
\nn
\eea
where $a\equiv 2 m_{\rm DM}/(x\sqrt{s})<1$, $\hat k=(0,\sin\theta_0,\cos\theta_0)$, and 
$\theta_0$ is the polar angle of $\hat k$ with respect to the beam line, in the c.o.m. frame.
The subscript $_0$ refers to quantities in the c.o.m. frame.
The three-momentum conservation fixes the angle $\theta_{0\,3j}$
 between $\hat p_3$ and $\hat k$:
$ \cos\theta_{0\,3j}=(\mathbf{p}_4^2-\mathbf{k}^2-\mathbf{p}_3^2)/{2|\mathbf{k}||\mathbf{p}_3|}$.
With these expressions, the squared amplitudes (\ref{MsquaredUV})-(\ref{Msquaredeff}) simplify to
\bea
\sum\overline{|{\cal M}_{\rm UV}|^2}&=&
\frac{8}{3}\,x^2s\frac{ Q_q^2e^2 g_q^2 g_\chi^2}{[x^2 s(1-z_0)-M^2]^2}
\frac{[1-z_0-a^2][1+(1-z_0)^2]}{z_0^2\sin^2\theta_0}\,,\\
\sum\overline{|{\cal M}_{\rm eff}|^2}&=&
\frac{8}{3}\,x^2s\frac{ Q_q^2e^2 }{\Lambda^4}
\frac{[1-z_0-a^2][1+(1-z_0)^2]}{z_0^2\sin^2\theta_0}\,,
\eea
which do not depend on  angles other than $\theta_0$, so we can simply
integrate the phase space over  the azimuth of $\mathbf{k}$ and over $\theta_{0\,3j}, \phi_{0\,3j}$
\be
\de \Phi_3=\frac{1}{(4\pi)^3}\de E_3\, \de k\,\de\cos\theta_0
=\frac{1}{(4\pi)^3}\frac{x^2 s}{4}\de z_0\,  \de y_0\,\de\cos\theta_0\,.
\ee
The kinematical domains of the variables $y_0,z_0$ are
\bea
\frac{z_0}{2}\left(1-\sqrt{\frac{1-z_0-a^2}{1-z_0}}\right)\leq
&y_0& \leq \frac{z_0}{2}\left(1+\sqrt{\frac{1-z_0-a^2}{1-z_0}}\right)\, ,\\
0\leq &z_0&\leq 1-a^2\,.
\eea
Finally, we get the differential cross sections with respect to the energy and angle of the emitted photon, 
in the c.o.m. frame
\bea
\left.\frac{\de^2\hat\sigma}{\de z_0\de\cos\theta_0}\right\vert_{\rm UV}
&=&
\frac{1}{3}\frac{Q_q^2\alpha g_q^2 g_\chi^2}{16\pi^2}\frac{1}{x^2 s}\frac{1}{\left[(1-z_0)-\frac{M^2}{x^2 s}\right]^2}
\frac{\left[1-z_0-\frac{4 m_{\rm DM}^2}{x^2 s}\right]^{3/2}}{\sqrt{1-z_0}}\frac{[1+(1-z_0)^2]}{z_0\sin^2\theta_0}
\label{d2sigmaUV}\\
\left.\frac{\de^2\hat\sigma}{\de z_0\de\cos\theta_0}\right\vert_{\rm eff}
&=&
\frac{1}{3}\frac{Q_q^2\alpha}{16\pi^2}\frac{x^2 s}{\Lambda^4}
\frac{\left[1-z_0-\frac{4 m_{\rm DM}^2}{x^2 s}\right]^{3/2}}{\sqrt{1-z_0}}\frac{[1+(1-z_0)^2]}{z_0\sin^2\theta_0}\,,
\label{d2sigmaeff}
\eea
where $\alpha=e^2/(4\pi)$. Eq.~(\ref{d2sigmaeff})  agrees with the findings  in Refs.~\cite{Chae:2012bq, Dreiner:2012xm}, up to the factor of 1/9, as we are considering colored colliding particles.

To get the cross sections in the lab frame we perform a 
 boost along the $\hat z$-axis, accounting for generic parton momentum fractions $x_1, x_2$,
 as in Eq.~(\ref{momfractions}).
 The  relations between the quantities $z_0,\theta_0$
 in the c.o.m. frame the the analog ones $z, \theta$ in the lab frame are
\bea
z_0&=&z\frac{(x_1+x_2)^2+\cos\theta(x_2^2-x_1^2)}{4x_1x_2}\\
\sin^2\theta_0\, ,
&=&\frac{4x_1x_2}{[(x_1+x_2)+\cos\theta(x_2-x_1)]^2}\sin^2\theta\,,
\eea 
so that the cross section in the lab frame is simply
\be
\frac{\de^2\hat\sigma}{\de z\de\cos\theta}=
\frac{x_1+x_2}{x_1+x_2+\cos\theta(x_2-x_1)}
\left.\frac{\de^2\hat\sigma}{\de z_0\de\cos\theta_0}\right\vert_{
\begin{footnotesize}
\begin{array}{l}
z_0\to z_0(z)\\
\theta_0\to \theta_0(\theta)
\end{array}
\end{footnotesize}
}\,.
\label{boosting}
\ee
Expressing the energy of the photon in terms of the transverse momentum and rapidity, $k^0=p_{\rm T} \cosh\eta$, one finds
\be
 z= \frac{4p_{\rm T}\cosh\eta}{(x_1+x_2)\sqrt{s}}\,,\qquad
\cos\theta= \tanh\eta\,,
\ee
which allows to express the differential cross sections with 
 respect to the transverse momentum and pseudo-rapidity of the
emitted photon, 
\be
\frac{\de^2\hat\sigma}{\de p_{\rm T}\de\eta}=
\frac{4}{(x_1+x_2)\sqrt{s}\cosh\eta}
\frac{\de^2\hat\sigma}{\de z\de\cos\theta}\,. 
\label{boosting2}
\ee
Therefore, from Eqs.~(\ref{d2sigmaUV}), (\ref{d2sigmaeff}), (\ref{boosting}) and (\ref{boosting2}), 
we finally get the desired cross sections   in the lab frame
\bea
\left.\frac{\de^2\hat\sigma}{\de p_{\rm T}\de\eta}\right\vert_{\rm UV}
&=&
\frac{Q_q^2 \alpha g_q^2g_\chi^2}{48 \pi^2}\frac{1}{ x_1 x_2 s}\frac{1}{p_{\rm T}}
\frac{\left[1-f-\frac{4m_{\rm DM}^2}{x_1x_2 s}\right]^{3/2}}
{\left[1-f-\frac{M^2}{x_1x_2 s}\right]^2}
\frac{\left[1+\left(1-f\right)^2\right]}{\sqrt{1-f}}\,,
\label{d2sigmaUVlab0}\\
\left.\frac{\de^2\hat\sigma}{\de p_{\rm T}\de\eta}\right\vert_{\rm eff}
&=&\frac{Q_q^2 \alpha}{48 \pi^2}\frac{x_1x_2 s}{\Lambda^4}
\frac{1}{p_{\rm T}}
\frac{\left[1-f-\frac{4m_{\rm DM}^2}{x_1x_2 s}\right]^{3/2}\left[1+\left(1-f\right)^2\right]}{\sqrt{1-f}}\,,
\label{d2sigmaefflab0}
\eea
where we have defined
\be
f(p_{\rm T}, \eta_,x_1,x_2)\equiv
\frac{p_{\rm T}(x_1 e^{-\eta}+x_2 e^\eta)}{x_1 x_2\sqrt{ s}}\,.
\ee
For the emission of a gluon, rather than a photon, one simply replaces 
$Q_q^2\alpha \to (4/3)\, \alpha_s$ in Eqs.~(\ref{d2sigmaUVlab0})-(\ref{d2sigmaefflab0}). 
These expressions reproduce the ones reported in Eqs.~(\ref{d2sigmaefflab})-(\ref{d2sigmaUVlab}).

\end{document}